%% file: TowardsVolume_v2.tex
\documentclass[11pt]{article}
\pdfoutput=1

\usepackage{jheppub}
\usepackage[lofdepth,lotdepth]{subfig} 

\usepackage{import}

\usepackage{bigints}

\title{Towards the Amplituhedron Volume}
\author[1]{Livia Ferro,}\emailAdd{livia.ferro@lmu.de}
\author[2]{Tomasz \L ukowski,}\emailAdd{lukowski@maths.ox.ac.uk}
\author[1]{Andrea Orta,}\emailAdd{andrea.orta@lmu.de}
\author[1]{and Matteo Parisi}\emailAdd{matteo.parisi@campus.lmu.de}

\affiliation[1]{Arnold--Sommerfeld--Center for Theoretical Physics,\\ Ludwig--Maximilians--Universit\"at, \\Theresienstra\ss e 37, 80333 M\"unchen, Germany }
\affiliation[2]{ Mathematical Institute, University of Oxford,\\ Andrew Wiles Building, Radcliffe Observatory Quarter,\\ Woodstock Road, Oxford, OX2 6GG, UK}

\abstract{It has been recently conjectured that scattering amplitudes in planar $\mathcal{N}=4$ super Yang-Mills are given by the volume of the (dual) amplituhedron. In this paper we show some interesting connections between the tree-level amplituhedron and a special class of differential equations. In particular we demonstrate how the amplituhedron volume for NMHV amplitudes is determined by these differential equations. The new formulation allows for a straightforward geometric description, without any reference to triangulations. Finally we discuss possible implications for volumes related to generic N$^k$MHV amplitudes.}

\begin{document}
\begin{flushright}
{\small LMU-ASC 77/15}
\end{flushright}
\maketitle


\section{Introduction}
\label{introduction}

In recent years we have observed a remarkable revolution in our understanding of basic principles of quantum field theories. The change of perspective was inevitable after the discovery of the AdS/CFT correspondence \cite{Maldacena:1997re} and it was additionally boosted by Witten's twistor string theory \cite{Witten:2003nn}. The most relevant impact of the latter can be seen in the way we think about the scattering amplitudes. It led to various new techniques for evaluating them, where the traditional Feynman's approach is replaced by purely on-shell methods. In particular, these new methods avoid introducing off-shell redundancies and, more importantly, they make the simplicity and the symmetry of the final answer manifest. The prime example where the influence of both the AdS/CFT correspondence and twistor string theory is most visible is the maximally supersymmetric gauge theory in four dimensions -- $\mathcal{N}=4$ SYM. For this theory, the on-shell methods were developed even further and led to a purely geometric description of scattering amplitudes \cite{ArkaniHamed:2010gg,ArkaniHamed:2012nw}.  The emergent picture allows to calculate them, at least in the planar limit, as ``volumes'' of an object termed amplituhedron \cite{Arkani-Hamed:2013jha}. Even though clear in concept and simple in definition, the amplituhedron is still waiting to show its true power. One of the reasons is its high complexity: for a given number of scattering particles, and for a given order in perturbation theory, the amplituhedron defines a complicated region in a high dimensional space. Then, in order to evaluate the amplitude, we need to find a differential form which has logarithmic singularities on all boundaries of this region. At the moment, there is no compact formula describing this differential form and the usual way to perform calculations is to refer to triangulations of the amplituhedron, also not known in general. Thus we are left with a case by case study where our new diagrammatics involves the evaluation of ``volumes'' of amplituhedron cells instead of Feynman diagrams. There is, however, a case where we can take full advantage of the volume concept: the so-called next-to-MHV (NMHV) amplitudes at tree level. In this case the ``volume'' is the true volume of an object dual to the amplituhedron. Moreover, there exists a formula, proposed by Hodges, which gives a unifying prescription on how to think of NMHV scattering amplitudes in this way \cite{Hodges:2009hk}. Although he still referred to triangulations, we show in this paper that this is not necessary and there exists a version of his formula treating the amplituhedron as a single object, independently of its triangulations. For N$^k$MHV amplitudes with $k>1$ a similar volume formula is not known at the moment and in this paper we would like to suggest a possible direction where to look for it.

There is no doubt that in order to properly and efficiently describe any physical process it is crucial to find a suitable set of variables. This holds true for the scattering amplitudes in $\mathcal{N}=4$ SYM as well. From the perspective of the space-time Lagrangian, the spinor-helicity variables $(\lambda^\alpha,\tilde\lambda^{\dot\alpha})$ are the most suitable for the description of massless scattering in four dimensions. When supplemented by Grassmann-odd variables parametrizing the \mbox{R-symmetry}, they allow to describe scattering in $\mathcal{N}=4$ SYM in a very compact way in terms of superfields
\begin{equation} \label{superfield}
\Phi = G^+ + \eta^A\,\Gamma_A + \frac{1}{2} \eta^A\eta^B\,S_{AB} + \frac{1}{3!} \eta^A\eta^B\eta^C \epsilon_{ABCD}\,\bar\Gamma^D + \frac{1}{4!} \eta^A\eta^B\eta^C\eta^D \epsilon_{ABCD}\,G^-\,.
\end{equation}
Then each superamplitude $\mathcal A_n(\{\Phi_i\})=\mathcal A_n(\{\lambda_i,\tilde\lambda_i,\eta_i\})$ is labelled by the number of scattering particles $n$ and, when the MHV part is factored out
\begin{equation} \label{factorMHV}
\mathcal A_n(\{\lambda_i,\tilde\lambda_i,\eta_i\}) = \mathcal A_{n,\textrm{tree}}^{\textrm{MHV}}  \mathcal P_n(\{\lambda_i,\tilde\lambda_i,\eta_i\}) \quad,\quad \mathcal A_{n,\textrm{tree}}^{\textrm{MHV}} = \frac{\delta^4(\sum \lambda_i\tilde\lambda_i)\delta^8(\sum \lambda_i\eta_i)}{\langle 12\rangle\langle 23\rangle\cdots\langle n 1\rangle} \,,
\end{equation}
it admits the following decomposition in various helicity sectors 
\begin{equation}
\mathcal P_n = \mathcal P_n^{\textrm{MHV}} + \mathcal P_n^{\textrm{NMHV}} + \mathcal P_n^{\textrm{N$^2$MHV}} + \dots + \mathcal P_n^{\overline{\textrm{MHV}}}\,,
\end{equation}
where each $\mathcal P_n^{\textrm{N$^k$MHV}}$ is a monomial in the $\eta$'s of order $\mathcal O(\eta^{4k})$.
The spinor-helicity variables obscure, however, a lot of nice properties of scattering amplitudes. In particular,
the tree-level scattering amplitudes $\mathcal A_n^{\mathrm{tree}}$ are invariant under the superconformal symmetry, a fact which is not manifest in these variables. In order to improve on that, we can perform a Fourier transform of the $\lambda^\alpha$ variables and end up with Penrose (super-)twistor variables, that linearize the generators of the superconformal algebra $\mathfrak{psu}(2,2|4)$.
Moreover, in the planar sector, $\mathcal{N}=4$ SYM enjoys an even bigger symmetry. Upon introducing new dual variables, 
one can show that tree-level scattering amplitudes are invariant under another copy of the $\mathfrak{psu}(2,2|4)$ algebra \cite{Drummond:2008vq}. The interplay with the previously mentioned one gives rise to the celebrated infinite-dimensional Yangian symmetry \cite{Drummond:2009fd}, described by the algebra $Y(\mathfrak{psu}(2,2|4))$. 
The super-twistor variables associated to the dual space are called momentum super-twistors and linearize the action of the dual superconformal symmetry.
Both ordinary super-twistors and momentum super-twistors provide a perfect framework to describe scattering amplitudes in $\mathcal{N}=4$ SYM. In particular, they made possible a major progress after the authors of \cite{ArkaniHamed:2008gz} proposed a description of amplitudes based on Grassmannian integrals, manifestly Yangian invariant \cite{ArkaniHamed:2009vw, Drummond:2010qh, Drummond:2010uq}. Firstly advocated in super-twistor space \cite{ArkaniHamed:2009dn} and then in momentum super-twistor space \cite{Mason:2009qx}, Grassmannian integrals compute scattering amplitudes as integrals over the set of $k$-planes in an $n$-dimensional space, $G(k,n)$. The connection to the Grassmannian space opened up new ways of studying amplitudes and, in particular, showed new connections with combinatorics and geometry. It eventually led to the formulation of the amplituhedron volume conjecture mentioned above. The amplituhedron is defined in yet another space which can be derived from the momentum super-twistor space by bosonizing its Grassmann-odd coordinates. Surprisingly, although this new set of variables makes geometric properties of amplitudes manifest, it obscures some of the algebraic ones. In particular, it is not clear how to realize the Yangian symmetry directly in this space. From the perspective of this paper, this fact is one of the obstacles to overcome in order to be able to derive a volume formula for $k>1$.

When working in the bosonized momentum twistor space, the amplituhedron differential form is expressed in terms of a set of positive external data $Z_{i}^A$, $i=1,\ldots,n$, $A=1,\ldots,k+m$, and an auxiliary set of vectors $Y_\alpha^A$, $\alpha=1,\ldots,k$. Here, $n$ is the number of particles, $k$ is the next-to-MHV degree and $m$ is an even number, which for true scattering in four dimensions is four. 
The aforementioned Grassmannian integrals are taken over matrix elements labelled by the indices $\alpha$ and $i$, hence the number of integrations grows with the number of scattering particles.
On the other hand, based on the NMHV case, volume integrals are taken over another Grassmannian, whose coordinates are rather indexed by $\alpha$ and $A$; varying the number of particles only affects the domain of integration. We name the latter space the dual Grassmannian and claim that it gives a natural set of coordinates in terms of which we can write a volume formula for any $k$. These two auxiliary Grassmannian manifolds arise naturally when studying a particular set of differential equations, called Capelli differential equations.

In this paper we address the question of finding the volume directly in the dual Grassmannian. We analyze the symmetries of the amplituhedron volume form and derive the differential equations it satisfies. Starting from these equations we are able to derive a novel dual space representation for the NMHV case and restrict its form for $k>1$. 
The paper is organized as follows. In section \ref{Sec:amplituhedron} we review the main relevant notions and specifically the Grassmannian integral and the amplituhedron for tree-level scattering amplitudes. In section \ref{Sec:Capelli} we define a set of differential equations obeyed by the amplituhedron volume. It consists of invariance and scaling properties, as well as the Capelli differential equations. A solution to these equations provides a novel formula computing the volume for NMHV tree-level scattering amplitudes, which we extensively check in the $m=2$ and $m=4$ cases. As a byproduct, we get a natural prescription to perform the original Grassmannian integrals. We also restrict the possible form of the volume for $k>1$. We conclude by pointing out the relation between our new formula and the deformed amplitudes.


\section{The amplituhedron}
\label{Sec:amplituhedron}

As already mentioned in the Introduction, tree-level scattering amplitudes can be calculated using Grassmannian integrals.
In terms of momentum super-twistors $\mathcal{Z}_i^{\mathcal A} = (\lambda_i^\alpha,\tilde\mu_i^{\dot\alpha},\chi_i^{\mathsf A})$, we have
\begin{equation} \label{Grassint}
\mathcal A_{n,k}^{\text{tree}} = \frac{1}{\textrm{Vol}(GL(k))}\bigintsss  \frac{d^{k\cdot n}\,c_{\alpha i}}{(12 \dots k)(23 \dots k+1)\dots(n1 \dots k-1)} \prod_{\alpha=1}^k \delta^{4|4}\left(\sum_{i=1}^n c_{\alpha i}\mathcal{Z}_i\right),
\end{equation}
where the integral is evaluated along a suitable complex contour and generates tree-level N$^k$MHV $n$-point amplitudes.
The integration is performed over the complex entries of a matrix $C$ spanning the Grassmannian $G(k,n)$, i.e.~the set of $k$-planes in the $n$-dimensional complex space. The volume factor removes the $GL(k)$  redundancy and $(i \dots i+k-1)$ denotes the $i$-th consecutive maximal minor of $C$.
 The contributions coming from this integral can be matched with certain on-shell diagrams, objects naturally appearing in positroid stratifications of Grassmannians. This identification led the authors of \cite{ArkaniHamed:2012nw} to relate scattering amplitudes with the positive Grassmannian: residues of the integral \eqref{Grassint} are in one-to-one correspondence with cells of $G_+(k,n)$, i.e.~the part of the Grassmannian $G(k,n)$ with all $k \times k$ ordered minors positive. This reduces the problem of computing amplitudes to that of combining cells of the positive Grassmannian. Pursuing the geometrization of the problem further led to the definition of the amplituhedron.

The amplituhedron is a new mathematical object whose volume is conjectured to compute the scattering amplitudes of planar $\mathcal{N}=4$ SYM. 
In the following we focus on the tree amplituhedron, i.e.~the object computing tree-level amplitudes. 
To define it one has to choose positive external data, namely $Z^A_i \in M_+(n,m+k)$, where $M_+(n,m+k)$ is the set of $ n \times (m+k)$ real matrices whose ordered\footnote{Note that ordered minors are not necessarily \emph{consecutive} minors.} maximal minors are positive:
\begin{equation}\label{positivity}
\langle Z_{i_1}\dots Z_{i_{m+k}} \rangle > 0 \qquad,\qquad \textrm{where}\qquad
\begin{cases} i_1,\dots,i_{m+k} = 1,\dots,n \\
i_1<\ldots<i_{m+k}\, 
\end{cases}.
\end{equation}
The tree amplituhedron is now the space
\begin{equation}\label{defamplituhedron}
\mathfrak{A}^{\mathrm{tree}}_{n,k;m}[Z] := \bigg\{ Y \in G(k,m+k) \;\;:\;\; Y = C \cdot Z \quad,\quad C \in G_+(k,n) \bigg\} \, ,
\end{equation}
where
\begin{equation} \label{Y=CZ}
Y^A_{\alpha}=\sum_i c_{\alpha i} Z^A_i ~.
\end{equation}
Therefore, the tree amplituhedron is a subspace of the Grassmannian $G(k,m+k)$ determined by positive linear combinations of positive external data. 
One can define an $(m\cdot k)$-dimensional canonical (top) form $\tilde{\Omega}_{n,k;m}(Y,Z)$ on this space, determined by the requirement that it has logarithmic singularities on all its boundaries. 
One way to obtain such form is to triangulate $\mathfrak{A}^{\mathrm{tree}}_{n,k;m}[Z]$, i.e.~to find a set of $(m \cdot k) $-dimensional cells of $G_+(k,n)$ such that the corresponding regions on the amplituhedron are non-overlapping and cover it completely.   
Once one has a triangulation $\mathcal{T}=\{\Gamma_a \}$, the canonical form associated to each cell $\Gamma$ is
\begin{equation}
\tilde{\Omega}_{n,k;m}^{\Gamma}(\beta^\Gamma)= \prod_{r=1}^{m \cdot k} \frac{\mbox{d} \beta^{\Gamma}_r}{\beta^{\Gamma}_r} \, ,
\end{equation}
where $(\beta^{\Gamma}_1,\dots ,\beta^{\Gamma}_{m \cdot k})$ are positive real coordinates on the cell.
One can now express that form in a coordinate-independent way solving for the $ \beta^{\Gamma}_r$ variables in terms of the  $Z_i^A$ and $Y_\alpha^A$ ones using equation \eqref{Y=CZ}, where the $c_{\alpha i}$'s are to be thought of as functions of the local coordinates $\beta_r^\Gamma$'s. Finally, the canonical form on the full
amplituhedron is the sum of the forms associated with each cell
\begin{align}
\tilde{\Omega}_{n,k;m}(Y,Z)= \sum_{\Gamma \in \mathcal{T}} \tilde{\Omega}_{n,k;m}^{\Gamma}(Y,Z) \,.
\end{align}
To build a connection with the tree-level super-amplitudes one relates the bosonic variables $Z_i^A$ with the momentum super-twistors. The components of the former consist of the momentum twistor variables $z_i^a:=(\lambda_i^\alpha,\tilde\mu_i^{\dot\alpha}) $ together with the bosonized version of the fermionic variables $\chi_i^{\mathsf A}$: 
\begin{equation}\label{bosonizedZ}
Z_i^A =
\left (
  \begin{tabular}{c}
  $z_i^a$ \\
  $ \phi_1^{\mathsf A}\;\chi_{i\mathsf A}$ \\
   \vdots\\
   $ \phi_k^{\mathsf A}\;\chi_{i\mathsf A}$
  \end{tabular}
\right ) \qquad,\qquad A=1,\dots, k+m, \quad a,\mathsf A =1, \ldots, m\,,
\end{equation}
with the $\phi_{\alpha}^{\mathsf A}$'s auxiliary Grassmann parameters used to bosonize the $\chi_i^{\mathsf A}$'s.
The tree-level ``amplitude'' for generic $m$ is then calculated by integrating the canonical form in the following way:
\begin{equation}\label{amplitoampli}
\mathcal{A}_{n,k}^{\mathrm{tree}}(\mathcal{Z})= \int d^{m\cdot k}\,\phi \int  \delta^{m\cdot k}(Y;Y^*) \, {\tilde{\Omega}}_{n,k;m}(Y,Z)\,,
\end{equation}
where the projective $\delta$-function 
\begin{equation}
\label{Ystar}
\delta^{m\cdot k}(Y;Y^*) = \int d^{k\cdot k}\,\rho_{\alpha}^{~\beta} ~ (\det\rho)^m \delta^{k\cdot (k+m)}(Y_{\alpha}^A - \rho_\alpha^\beta Y^{*A}_\beta)
\end{equation}
localizes the canonical form on the reference point
\begin{equation}
\qquad Y^* =
\left (
  \begin{tabular}{c}
  $\mathbb{O}_{m\times k}$ \\
   - - -  \\
   $\mathbb{I}_{k\times k}$
  \end{tabular}
\right ) \,.
\end{equation}
In order to be consistent with the positivity conditions \eqref{positivity}, at the beginning one has to regard $Z^A_i$ as real numbers. Only after the localization of the canonical form, just before the integration over $\phi^{\mathsf A}_\alpha$, the last $k$ components of $Z^A_i$ have to be analytically continued to composite Grassmann variables $\phi_\alpha\cdot\chi_i$.

Triangulating the amplituhedron to derive the canonical form $\tilde\Omega_{n,k;m}$ is not the only available option. The authors of \cite{Arkani-Hamed:2014dca} suggested another method based on a more invariant approach to the positive geometry without referring to any triangulation. We point out that there exists yet another way to construct $\tilde\Omega_{n,k;m}$.
First, let us rewrite
\begin{equation} 
\label{Omeganotilde}
{\tilde{\Omega}}_{n,k;m}(Y,Z)=\prod_{\alpha=1}^k \langle Y_1\cdots Y_k \,d^m Y_{\alpha} \rangle \,{\Omega}_{n,k}^{(m)}(Y,Z) \,,
\end{equation}
where $\prod_{\alpha=1}^k \langle Y_1\cdots Y_k \,d^m Y_{\alpha} \rangle$ is the integration measure on $G(k,m+k)$ space.
Then, we can give an integral representation of ${\Omega}_{n,k}^{(m)}(Y,Z)$ 
\begin{equation} \label{omegaintegral}
\Omega_{n,k}^{(m)}(Y,Z) = \int \frac{d^{k \cdot n}\,c_{\alpha i}}{(1 2\ldots k) (2 3\ldots k+1)\ldots(n 1 \ldots k-1)} \prod_{\alpha=1}^k \delta^{m+k}(Y_{\alpha} - \sum_i c_{\alpha i}  Z_i).
\end{equation}
The integral is taken over a suitable contour, in full analogy with the Grassmannian integral \eqref{Grassint}. Each residue corresponds to a cell of the tree amplituhedron and, in order to get the proper expression for $\Omega^{(m)}_{n,k}$, we need to take an appropriate sum of them. 

Integral \eqref{omegaintegral} will be the starting point for our later derivation. In particular, we begin by considering its properties and symmetries in the case of generic $m$, $n$ and $k$. This allows us to write down a set of differential equations satisfied by $\Omega^{(m)}_{n,k}$. We subsequently solve this system for $k=1$ and restrict the form of possible solutions for higher $k$. In all cases the answer admits an integral representation over the Grassmannian $G(k,m+k)$. We term this space the dual Grassmannian and stress that it does not depend on the value of $n$, in contrast to \eqref{omegaintegral}. 


\section{Capelli differential equations and volume}\label{Sec:Capelli}

\subsection{Properties of the volume}

As we have already mentioned, the tree-level Grassmannian integrals \eqref{Grassint} defined in momentum twistor space possess a lot of interesting properties. In particular, they are superconformally and dual-superconformally invariant or, equivalently, Yangian invariant. As was shown in \cite{Drummond:2010uq,Korchemsky:2010ut} these symmetries uniquely determine, up to the contour of integration, the form of Grassmannian integrals and in particular fix their measure to be the inverse of the product of consecutive cyclic minors\footnote{This statement is true for Yangian generators with trivial local level-one generators which are relevant for scattering amplitudes. For a discussion on possible deformations of amplitudes and Yangian generators see \cite{Ferro:2012xw,Ferro:2013dga}.}. In this paper we aim at finding the formula for the amplituhedron volume defined in the bosonized momentum twistor space. It is then an interesting question to ask whether it is also possible to determine its form directly starting from symmetries. The answer we provide is positive, at least for the NMHV amplitudes. For N$^k$MHV amplitudes with $k\geq2$, however, known symmetries of the amplituhedron are not sufficient to completely fix the expression for the volume. In particular, it is not clear how to realize the Yangian symmetry directly in the bosonized momentum twistor space, preventing us from repeating the derivation yielding the Grassmannian measure in \eqref{Grassint}. Despite this obstacle, let us study the symmetries of the formal integral \eqref{omegaintegral} and use them to derive a formula for the volume.

First of all, the integral \eqref{omegaintegral} is $GL(m+k)$ covariant
\begin{equation}\label{invariancediff}
\sum_{a=1}^{n+k} W_a^A\frac{\partial}{\partial W_a^B}\,\Omega_{n,k}^{(m)}(Y,Z)=-k \,\delta^A_B\; \Omega_{n,k}^{(m)}(Y,Z) \,,
\end{equation}  
where we have defined 
\begin{equation}
W_a^A := 
\begin{cases}
Y_a^A  \;&,\quad a = 1,\dots,k , \\
Z^A_{a-k} \;&,\quad a = k+1,\dots,k+n .
\end{cases}
\end{equation}
This statement is analogous to the level-zero Yangian invariance of the Grassmannian formula \eqref{Grassint}. Moreover, $\Omega^{(m)}_{n,k}$ is invariant under rescaling of the variables $Z^A_i$
\begin{equation}\label{scalingZdiff}
\sum_{A=1}^{m+k} Z_{i}^{A}\frac{\partial}{\partial Z_{i}^A}\,\Omega^{(m)}_{n,k}(Y,Z)=0\,,\qquad \text{for }i=1,\ldots,n\,,
\end{equation}
and is a $GL(k)$-covariant homogeneous function of degree $-(m+k)$ with respect to the $Y_\alpha^A$ variables
\begin{equation}\label{scalingYdiff}
\sum_{A=1}^{m+k} Y_{\alpha}^{A}\frac{\partial}{\partial Y_{\beta}^A}\,\Omega^{(m)}_{n,k}(Y,Z)=-(m+k)\, \delta^\beta_\alpha\, \Omega^{(m)}_{n,k}(Y,Z),\qquad \text{for }\alpha,\beta=1,\ldots,k.
\end{equation}
Both scaling and level-zero Yangian invariance were ingredients which allowed to determine the Grassmannian measure. However, the knowledge of bilinear level-one Yangian generators was necessary to get a unique answer. Unfortunately at the moment we do not know what their form would be in the bosonized momentum twistor space beyond the $k=1$ case\footnote{It is not even clear whether the amplituhedron measure is Yangian invariant since, in order to derive the Yangian invariant formula \eqref{Grassint} from it, we project out many terms when the last $k$ components of $Z^A_i$ are taken to be composite Grassmann variables.}. Nevertheless, it is easy to verify that $\Omega_{n,k}^{(m)}(Y,Z)$ satisfies other higher-order differential equations: for every $(k+1)\times (k+1)$ minor of the matrix composed of derivatives $\frac{\partial}{\partial W^A_a}$ one can check that
\begin{equation}\label{Capelli}
\det\mathop{\left(\frac{\partial}{\partial W^{A_\nu}_{a_\mu}}\right)_{1\leq \nu \leq k+1}}_{\hspace{1.7cm}1\leq \mu\leq k+1} \Omega_{n,k}^{(m)}(Y,Z) =0\,,
\end{equation}
for $1\leq A_1 \leq \ldots \leq A_{k+1}\leq m+k$ and $1\leq a_1 \leq \ldots \leq a_{k+1} \leq n+k$. 
This type of determinant differential equations are usually referred to as the {\it Capelli differential equations}. In the case at hand, we consider the set of all possible Capelli differential equations defined on the Grassmannian $G(m+k,n+k)$. Interestingly, the Capelli equations \eqref{Capelli} together with the invariance property \eqref{invariancediff} and scaling properties of the form \eqref{scalingZdiff} and \eqref{scalingYdiff} were studied independently in the mathematical literature in various contexts. The most understood case is $k=1$, corresponding to the NMHV amplitudes, which leads to the definition of the so-called GKZ hypergeometric function (on Grassmannians).

In order to establish a connection with the known mathematical literature, we first rewrite the invariance and scaling conditions in their global form. For given $m$, $k$ and $n$ we want to find a function $\Omega^{(m)}_{n,k}(Y^A_{\alpha}, Z_i^A)$ with $Y^A_{\alpha} \in M(k,m+k)$ and $Z_i^A \in M(n,m+k)$ satisfying
\begin{itemize}
\item
$GL(m+k)$ right covariance: 
\begin{equation}\label{invarianceglobal}  
\Omega^{(m)}_{n,k}( Y \cdot g,  Z \cdot g)= \frac{1}{(\det g)^{k}} \,\Omega^{(m)}_{n,k}(Y,Z)\,,
\end{equation}
for $g \in GL(m+k)$, where by the right multiplication we mean $( W \cdot g)^{A}_a=\sum_{B}W^B_a g^{\, A}_{B}$.

\item
Scaling, i.e. $GL(k)_+\otimes GL(1)_+ \otimes\ldots \otimes GL(1)_+$ left covariance:
\begin{equation}\label{scalingglobal}
 \Omega^{(m)}_{n,k}(h\cdot Y ,\lambda\cdot Z) = \frac{1}{(\det h)^{m+k}}\, \Omega^{(m)}_{n,k}(Y,Z)\,,
\end{equation}
for $h \in GL(k)_+$ and $\lambda=(\lambda_1,\dots ,\lambda_n)\in GL(1)_+ \otimes\ldots \otimes GL(1)_+$, where we restricted all possible transformations to be elements of the identity component of linear groups, namely $GL(l)_+=\{ h\in GL(l):\det h>0\}$. The condition \eqref{scalingglobal} takes into account both conditions \eqref{scalingZdiff} and \eqref{scalingYdiff}.
 
\item Capelli differential equations on the Grassmannian $G(m+k,n+k)$ defined as in formula \eqref{Capelli}.
\end{itemize}

Functions satisfying the conditions \eqref{Capelli}, \eqref{invarianceglobal} and \eqref{scalingglobal} for the $k=1$ case were studied intensively by the school of Gelfand \cite{MR841131} and also by Aomoto \cite{Aomoto:1975, Aomoto}. In the context of scattering amplitudes in $\mathcal{N}=4$ SYM, their relevance was suggested in \cite{Ferro:2014gca}. For the $k=1$ case, the general solution of the above problem was given in \cite{GelfandGraev} and we will present it in the following section. It gives the correct result for $\Omega^{(m)}_{n,1}$ as an integral over the Grassmannian $G(1,m+1)$. As we will see below, this integral calculates the volume of a simplex in the projective space $G(1,m+1)=\mathbb{RP}^m$ and can be compared to the volume formula proposed by Hodges \cite{Hodges:2009hk}. An important advantage with respect to the latter is that it can be evaluated without referring to any triangulation of the simplex. For higher $k$ the problem was studied for example in \cite{Oshima:1995capelliidentities}, but to our knowledge a general solution suitable for the scaling properties \eqref{scalingglobal} is not known. 

We look for the solution to \eqref{Capelli}, \eqref{invarianceglobal} and \eqref{scalingglobal} written in the Fourier space 
\begin{equation}\label{OmegaFourier}
\Omega^{(m)}_{n,k}(Y,Z)= \int d\mu(t_A^{\alpha},\tilde t_A^i) \,e^{i\, t_A^{\alpha}\,Y^A_{\alpha}  + i\, \tilde t_A^i \,Z_i^A } f(t_A^{\alpha},\tilde t_A^i)\,,
\end{equation}
where the variables $t^\alpha_A$ and $\tilde t^{i}_A$ are Fourier conjugate to $Y^A_\alpha$ and $Z^A_i$, respectively. Here the flat measure $ d\mu(t_A^{\alpha},\tilde t_A^i)$ is both $GL(m+k)$ and $GL(k)$ covariant and $f(t_A^{\alpha},\tilde t_A^i)$ is a generalized function defined on the product of two matrix spaces: 
\begin{equation}
f:M(m+k,k)\times M(m+k,n)\to \mathbb{R}\,.
\end{equation}
In the following section we present the derivation for the $k=1$ case to emphasize our assumptions and prepare for the study of higher values of $k$.


\subsection{Solution for the \texorpdfstring{$k=1$}{} case }

For the $k=1$ case, relevant for NMHV amplitudes, the index $\alpha$ can take just one value and \eqref{OmegaFourier} reduces to
\begin{equation}\label{Omegak1}
\Omega^{(m)}_{n,1}(Y,Z)= \int d\mu(t_A,\tilde t_A^i)\, e^{i\, t_A\, Y^A + i \,\tilde t_A^i \,Z_i^A } f(t_A,\tilde t^i_A).
\end{equation}
Then the Capelli differential equations form a system of second-order differential equations. We distinguish two cases: both derivatives are with respect to $Z^A_i$ variables or one derivative is with respect to a $Z^A_i$ and another to $Y^A$. Explicitly, they read
\begin{align}
\left(\frac{\partial^2}{\partial Z_i^A \partial Z_j^B} - \frac{\partial^2}{\partial Z_i^B \partial Z_j^A}\right) \Omega^{(m)}_{n,1}  = 0\, \qquad \text{and} \qquad
\left(\frac{\partial^2}{\partial Y^A \partial Z_j^B} - \frac{\partial^2}{\partial Y^B \partial Z_j^A}\right) \Omega^{(m)}_{n,1} = 0\,.
\end{align}
When applied to the formula \eqref{Omegak1}, they can be translated into the following equations in Fourier variables:
\begin{align}\label{k1tttt}
\tilde t^i_A \tilde t^j_B  - \tilde t^i_B \tilde t^j_A =0 \qquad \text{and} \qquad
t_A \tilde t^i_B  - t_B \tilde t^i_A =0\,.
\end{align}
It is immediate to verify that 
\begin{equation}
\tilde t_A^i = - \tilde s^i s_A , \qquad t_A = s_A\,,
\end{equation} 
is a solution of \eqref{k1tttt} for any $n$ and transforms the Fourier integral \eqref{Omegak1} into
\begin{equation}\label{k1preFourier}
\Omega^{(m)}_{n,1}(Y,Z)= \int ds_A \,d\tilde s^i\, e^{i\, s_A\,Y^A - i\, s_A \,Z_i^A \,\tilde s^i } F(s,\tilde s)\,,
\end{equation}
where we integrate over the space $\mathbb{R}^{m+1}\times \mathbb{R}^{n}$. Therefore, from the perspective of Capelli differential equations the most convenient and natural  variables are not the Fourier ones, but rather the ones we call $s_A$ and $\tilde s^i$. Every $\tilde s^i$ can be identified with the corresponding $c_i$ in the integral \eqref{omegaintegral} for $k=1$, while we will refer to $s_A$ as dual variables. From the definition of the amplituhedron we demand that \eqref{k1preFourier} localizes on
\begin{equation}
Y^A=\tilde s^i Z_i^A\,,
\end{equation}
as in \eqref{omegaintegral}. This is only possible if the function $F(s,\tilde s)$ is independent of $s_A$: indeed, upon integration over $s_A$ in \eqref{k1preFourier}, we would end up with the desired $\delta$-function. Then we can write a representation of $\Omega^{(m)}_{n,1}(Y,Z)$ purely as an integral over dual variables
\begin{equation}
\Omega^{(m)}_{n,1}(Y,Z)= \int ds_A \, e^{i\, s_A\,Y^A } \tilde F(s_A Z^A_i )\,,
\end{equation}
where $\tilde F(s_A Z^A_i )$ is the Fourier transform of the function $F(\tilde s)$. Notice that the integrand depends on the external data only through the $n$ combinations $s_A Z^A_i$.

Let us observe that in the $k=1$ case the fact that $\Omega_{n,k}^{(m)}$ satisfies the Capelli differential equations and scaling properties directly implies that it is also invariant under the level-one Yangian generators of the form
\begin{equation}\label{Yangiangen}
\hat J_B^A=\sum_{a<b}\left(W_a^A \frac{\partial}{\partial W_a^C}W_b^C \frac{\partial}{\partial W_b^B}-(a\leftrightarrow b)\right)+ (m+1)Y^A\frac{\partial}{\partial Y^B}\,.
\end{equation} 
When supplemented by \eqref{invariancediff}, i.e.~the level-zero Yangian invariance condition, it implies the full Yangian invariance for $k=1$. This statement is however not true for $k>1$.

This ends the study of the Capelli differential equations. 
 Now we need to supplement it by the invariance and scaling properties, which will constrain possible forms of the function $\tilde F(s_A Z^A_i )$. After careful analysis we find that it has to be a homogeneous (generalized) function of degree zero in each of its variables.
The space of homogeneous generalized functions is a well studied one. Following \cite{gelʹfand1964generalized}, we find that for each integer number $l$ there are exactly two independent homogeneous generalized functions of degree $l$.  For $l=0$, one can pick as a basis the Heaviside step functions $\theta(x)$ and $\theta(-x)$.
This yields the general solution
\begin{equation}\label{k1withconsts}
\Omega^{(m)}_{n,1}(Y,Z)= \int ds_A \, e^{i\,s_A\, Y^A  }  \prod_i \left(C_i \,\theta\left(s_A Z_i^A\right)+D_i\,\theta\left(-s_A Z_i^A\right)\right)\,,
\end{equation}
where $C_i$ and $D_i$ are arbitrary complex numbers. The existence of various solutions can be linked with the ambiguity in choosing the integration contour of the Grassmannian integral. By direct calculation we find that the solutions relevant for scattering amplitudes are the ones with either all $D_i=0$ or all $C_i=0$. In the first case we end up with 
\begin{equation}\label{volume1}
\Omega^{(m)}_{n,1}(Y,Z)=\frac{1}{i^{m+1}} \int ds_A \, e^{i \,s_A\,Y^A } \prod_i  \theta(s_A \,Z^A_i)\,.
\end{equation}
As we will show shortly, this is the correct formula for the volume. Before we do it, let us rewrite \eqref{volume1} in a way that resembles the formula found by Hodges. First, let us observe that \eqref{volume1} is $GL(m+1)$ covariant and use this to fix $m+1$ of the $Z_i$'s to form an identity matrix, namely, $\{Z_1,\dots,Z_{m+1}\}=\mathbb I_{m+1}$. Then 
 \begin{equation}\label{volume2}
\Omega^{(m)}_{n,1}(Y,Z)= \frac{1}{i^{m+1}}\int\displaylimits_0^{+\infty} \left(\prod_{A=1}^{m+1} ds_A\right) \,  e^{i\, s_A\, Y^A} \prod_{i=m+2}^n  \theta(s_A\, Z^A_i)\,,
\end{equation}
where we used $m+1$ of the $\theta$-functions to restrict the domain of integration. Furthermore, we can perform a change of variables $s \to s'$ such that
\begin{equation}
s_1 = s'_1\,,\qquad\qquad s_A = s'_1s'_A\,, \,\,\text{ for }\,\, A=2,\ldots,m+1\,,
\end{equation}
and compute the integral over $s'_1$ explicitly, to end up with 
\begin{eqnarray}\label{volume3}
\Omega^{(m)}_{n,1}= \int\displaylimits_0^{+\infty} \left(\prod_{A=2}^{m+1}ds_A\right) \frac{m!}{ \left(s\cdot Y\right)^{m+1}} \prod_{i=m+2}^n \theta\left(s\cdot Z_i\right) \,.
\end{eqnarray}
Here we introduced the compact notation $s \cdot W_a  := W_a^1 + s_2\,W_a^2  +\ldots+  s_{1+m} W_a^{1+m} $, where $W_a$ can be either $Y$ or one of the $Z_i$'s. Formula \eqref{volume3} is the most important formula of this section! Let us remark that this integral is taken over the $m$-dimensional real projective space $\mathbb{RP}^m$. A few comments are in order here. First of all, for positive external data and for $Y^A$ inside the amplituhedron, namely \eqref{defamplituhedron}, this integral is finite for any number of points $n$. It follows from the fact that in this case the poles of the integrand always lie  outside the integration region. Additionally, the behaviour at infinity guarantees the convergence. 
Moreover, one can compare the integral \eqref{volume3} with the one found in \cite{Hodges:2009hk} by Hodges. Then the elements of the projective space over which we integrate can be identified with the elements of dual momentum twistor space. We prefer to think about them rather as elements of the dual Grassmannian, as we explained in the introduction. This intuition has a natural generalization to higher $k$.

Before we show in detail how the formula \eqref{volume3} evaluates the volume of the amplituhedron, we comment on the relation of our formula to the formal expression \eqref{omegaintegral}. Let us consider again the integral \eqref{volume1},  write the Fourier representation of $\theta$-functions and, subsequently, integrate over all $s_A$:
\begin{align}
\Omega^{(m)}_{n,1}(Y,Z) 
&=\frac{i^{n-m-1}}{(2\pi)^n}  \int ds_A\, d\tilde s^i \, e^{i\,s_A\, Y^A - i\, s_A\,Z_i^A \,\tilde s^i } \prod_i \frac{1}{\tilde s^i + i \epsilon^i} \;= \notag\\
&=\frac{1}{(-2\pi i)^{n-m-1}}\int \prod_i \frac{d\tilde s^i}{\tilde s^i+i\epsilon^i}\, \delta^{m+1}\left(Y^A - \tilde s^i Z^A_i \right)\quad\,,
\end{align}
with all $\epsilon^i>0$. Here we integrate all variables over the real line, whereas \eqref{omegaintegral} was computed along some complex contour evaluating the proper sum of residues. The $i\epsilon$-prescription  with all $\epsilon^i$ positive was already advocated by Arkani-Hamed, see e.g.~\cite{Nima:2014}, and from our discussion we see that it has a natural origin in the dual space.

Let us spend a few words on the general structure of formula \eqref{volume3} and describe how to deal with the $\theta$-functions constraints. 
First of all, the integrand depends on the number of particles only through the $\theta$-functions shaping a domain of integration where no singularities are present. Indeed, 
\begin{equation}\label{nonsingintegrand}
s \cdot Y = s \cdot (c_i Z_i) = c_i \;(s \cdot Z_i) > 0\,,
\end{equation}
since $s \cdot Z_i > 0$ and $Y$ is inside the amplituhedron, i.e.~$c_i>0$. Furthermore, positivity of the external data implies that the domain is convex. 
Recall that the $GL(m+1)$ covariance of the integral \eqref{volume1} allows us to fix $m+1$ variables $\{Z_1,\dots,Z_{m+1}\}=\mathbb I_{m+1}$. From now on, we will work in this particular frame and only at the end of our calculation we will rewrite the results to be valid in general. For the $n$-point integral $\Omega^{(m)}_{n,1}$, let us denote the integration domain, defined by the $\theta$-functions in~\eqref{volume3}, by
\begin{equation}
\mathcal D^{(m)}_n:= \bigcap_{i=1}^n{\{s\cdot Z_i  > 0\}} = \bigcap_{i=m+2}^n{\{s\cdot Z_i  > 0\}} \cap \{s>0\}\,,
\end{equation}
where $\{s>0\}$ means that all $s_A$'s are positive, as dictated by the aforementioned fixing. 
We observe that
\begin{equation}\label{intdomains}
\Omega^{(m)}_{n,1} = m!\int_{\mathcal D^{(m)}_n} ds \; (s\cdot Y)^{-(m+1)} = \Omega^{(m)}_{n-1,1} - m!\int_{\mathcal D^{(m)}_{n-1}\cap\{s\cdot Z_n  < 0\}}ds \; (s\cdot Y)^{-(m+1)}\,,
\end{equation}
which will be extensively used later on. We also denote by $\ell_{Z_i}$ the \hbox{$(m-1)$\nobreakdash-dimensional} subspace $s\cdot Z_i = 0$ defined by the $\theta$-functions.

In the following subsections we present a detailed analysis of the volume integral for $k=1$. Although the scattering amplitudes in planar $\mathcal{N}=4$ SYM correspond to the case $m=4$, we find it advantageous to study formula \eqref{volume3} first in the two-dimensional toy model with $m=2$. In this case it takes the explicit form
\begin{eqnarray}\label{volumem2}
\Omega^{(2)}_{n,1}=\int\displaylimits_0^{+\infty} ds_2\int\displaylimits_0^{+\infty} ds_3 \,\frac{2}{ \left(s\cdot Y\right)^{3}} \prod_{i=4}^n \theta\left(s\cdot Z_i\right) .
\end{eqnarray}
Later on, we will discuss the formula for $m=4$
\begin{eqnarray}\label{volumem4}
\Omega^{(4)}_{n,1}= \int\displaylimits_0^{+\infty} ds_2\int\displaylimits_0^{+\infty} ds_3\int\displaylimits_0^{+\infty} ds_4\int\displaylimits_0^{+\infty} ds_5\, \frac{4!}{ \left(s\cdot Y\right)^{5}} \prod_{i=6}^n \theta\left(s\cdot Z_i\right) ,
\end{eqnarray}
 which is related to the physical scattering amplitudes.


\subsection{Volume in the \texorpdfstring{$m=2$}{} case}

For any number of points $n$, a possible representation of the volume $\Omega_{n,1}^{(2)}$ is given by \cite{ArkaniHamed:2010gg}
\begin{equation}\label{answer2}
\Omega_{n,1}^{(2)}=\sum_{i=2}^{n-1}\, [1\,i\,i+1]\,,
\end{equation}
where the R-invariants are defined as
\begin{equation}
[i\,j\,k]:=\frac{\langle i\,j\,k\rangle^2}{\langle Y \,i\,j\rangle \langle Y \,j\,k\rangle \langle Y \,k\,i\rangle}\,.
\end{equation}
We will verify that formula \eqref{volumem2} exactly reproduces this result. 

\begin{figure}[ht]
\centering
\def\svgwidth{5cm}
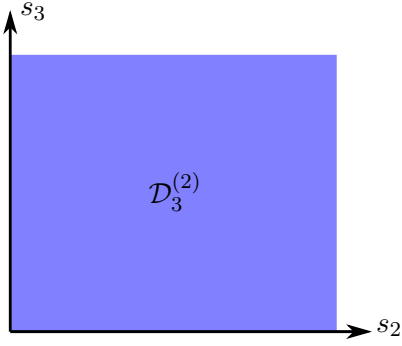
\caption{Domain of integration for three points.}\label{Fig.3points}
\end{figure}

For three points there are no $\theta$-functions and we have to evaluate the following integral
\begin{equation}
\Omega^{(2)}_{3,1} = \int\displaylimits_0^{+\infty} ds_2 \int\displaylimits_0^{+\infty}ds_3  \frac{2}{ \left(Y^1 + s_2 Y^2 +s_3 Y^3\right)^{3}}\,,
\end{equation}
where the region of integration is simply the positive quadrant in two dimensions, as in Fig.~\ref{Fig.3points}.
\noindent As discussed in \eqref{nonsingintegrand}, the integrand does not have poles inside $\mathcal D^{(2)}_3$. By performing the integral we simply find
\begin{equation}
\Omega^{(2)}_{3,1} = \frac{1}{Y^1 Y^2 Y^3}\,.
\end{equation}
There is a unique way to lift this formula to the case of generic $Z$'s by rewriting it in terms of $GL(3)$-invariant brackets with the proper scaling:
\begin{equation}
\Omega^{(2)}_{3,1} =\frac{\langle 123 \rangle^2}{\langle Y12 \rangle\langle Y23 \rangle\langle Y31 \rangle} = [123]\,.
\end{equation}
This agrees with the formula \eqref{answer2}.

\begin{figure}[ht]
\centering
\def\svgwidth{5cm}
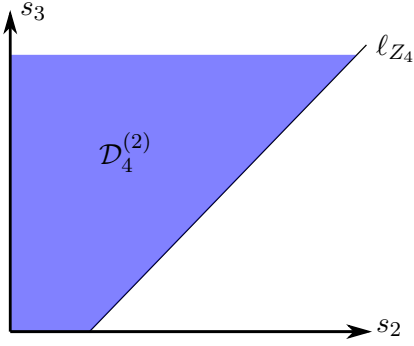
\caption{Domain of integration for four points.}
\label{Fig.4points}
\end{figure}

\begin{figure}[ht]
\centering
\subfloat[][Local triangulation]{\label{4pt_int}
\def\svgwidth{5cm}
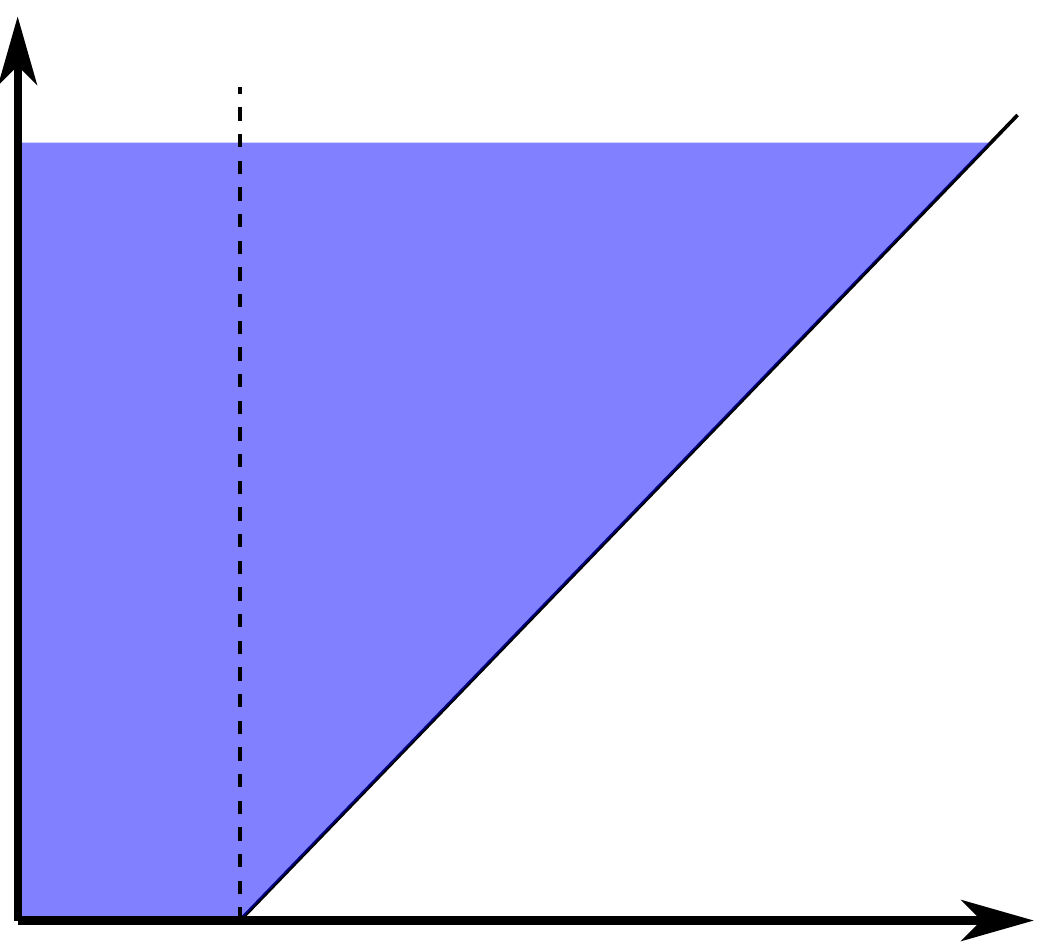} \qquad\qquad
\subfloat[][BCFW triangulation]{\label{4pt_ext}
\def\svgwidth{5cm}
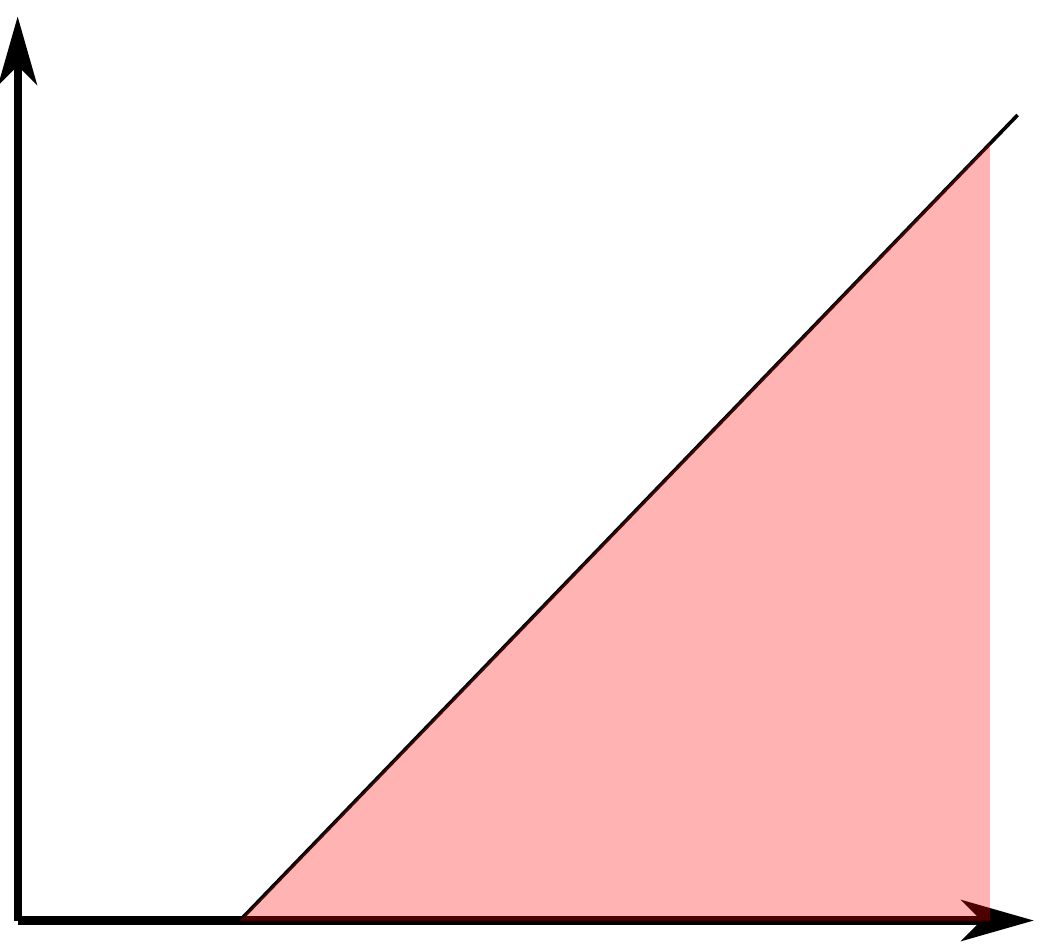}
\caption{Two ways of obtaining the four-point integral.}\label{Fig.4pointstwo}
\end{figure}

For four points the formula \eqref{volumem2} reads now
\begin{equation}
\label{I4}
\Omega^{(2)}_{4,1}  = \int\displaylimits_0^{+\infty} ds_2 \int\displaylimits_0^{+\infty}ds_3  \,\frac{2}{ \left(Y^1 + s_2 Y^2 +s_3 Y^3\right)^{3}}\;\theta\left(Z_4^1+s_2 Z_4^2 + s_3 Z^3_4\right) \,.
\end{equation}
Demanding positivity of the external data, we see that the components of $Z_4^A$ must satisfy $Z_4^1>0$, $Z_4^2<0$ and $Z_4^3>0$. Then, the $\theta$-function simply describes a half-plane in the $(s_2, s_3)$ plane above the line $\ell_{Z_4}: s\cdot Z_4=0$, which has positive slope and intersects the positive $s_2$ semi-axis, see Fig.~\ref{Fig.4points}.
It is straightforward to evaluate the integral \eqref{I4} explicitly, however, in order to make contact with results known in the literature, it is useful to think of the domain $\mathcal D^{(2)}_4$ in two different ways, depicted in Fig.~\ref{Fig.4pointstwo}.
On the one hand, we can split the integration region as in Fig.~\ref{4pt_int}, leading to the local (internal) triangulation \cite{ArkaniHamed:2010gg}
\begin{equation}
\Omega^{(2)}_{4,1}=\{ 3\}+\{ 4\}\,,
\end{equation}
with
\begin{equation}
\{i\} :=\frac{\langle 12 i\rangle \langle i-1 \,i\,i+1\rangle}{\langle Y 12\rangle \langle Y\,i-1\, i\rangle \langle Y\, i\, i+1\rangle}\,.
\end{equation}
Alternatively, we can obtain it as the difference of $\mathcal D^{(2)}_3$ with the region shown in Fig.~\ref{4pt_ext}: this choice produces an external triangulation, agreeing with the terms coming from BCFW recursion relations\footnote{In order to be able to perform the integral over the region in Fig.~\ref{4pt_ext} one needs to additionally assume that the integrand does not have any pole there, since it is not ensured by the geometry of the amplituhedron.}
\begin{equation}
\Omega^{(2)}_{4,1} = [123] + [134]\,.
\end{equation}

\begin{figure}[ht]
\centering
\subfloat[][Integration domain]{\label{Fig.m25points}
\def\svgwidth{4,4cm}
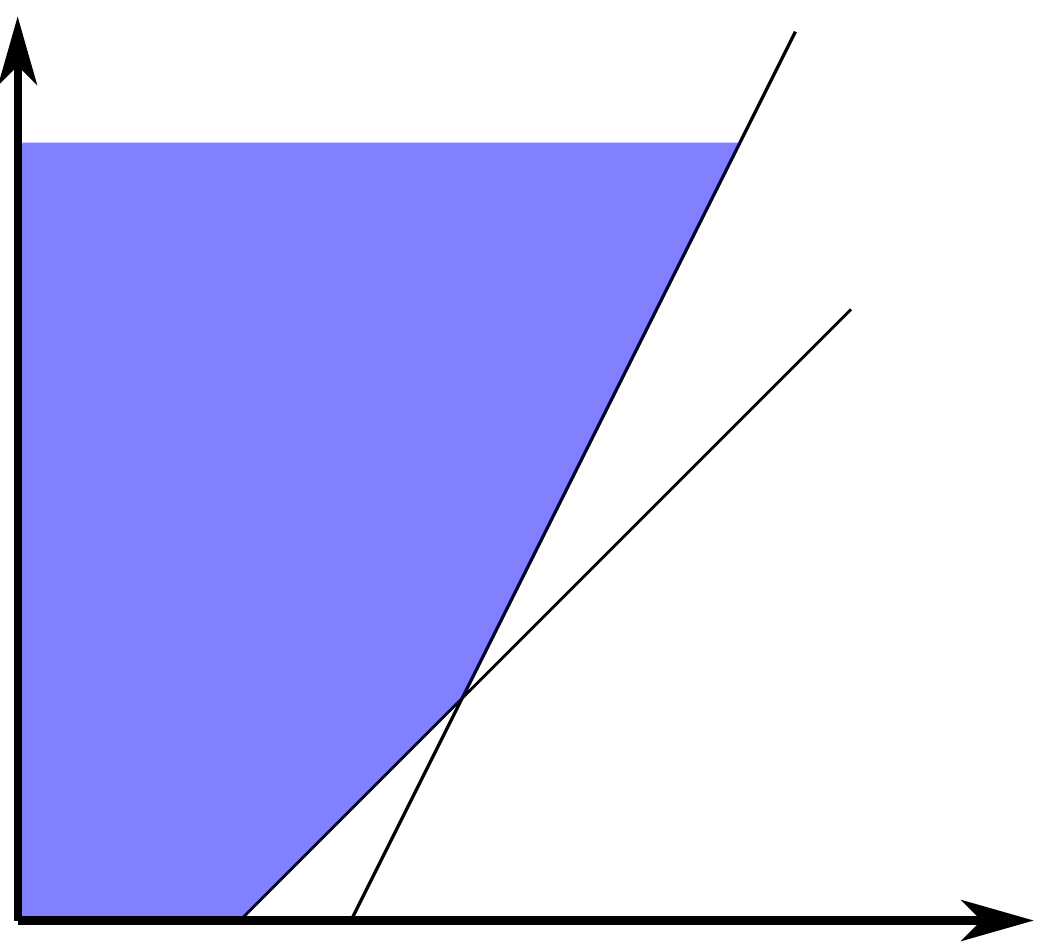} \qquad
\subfloat[][Local triangulation]{\label{5pt_int}
\def\svgwidth{4,4cm}
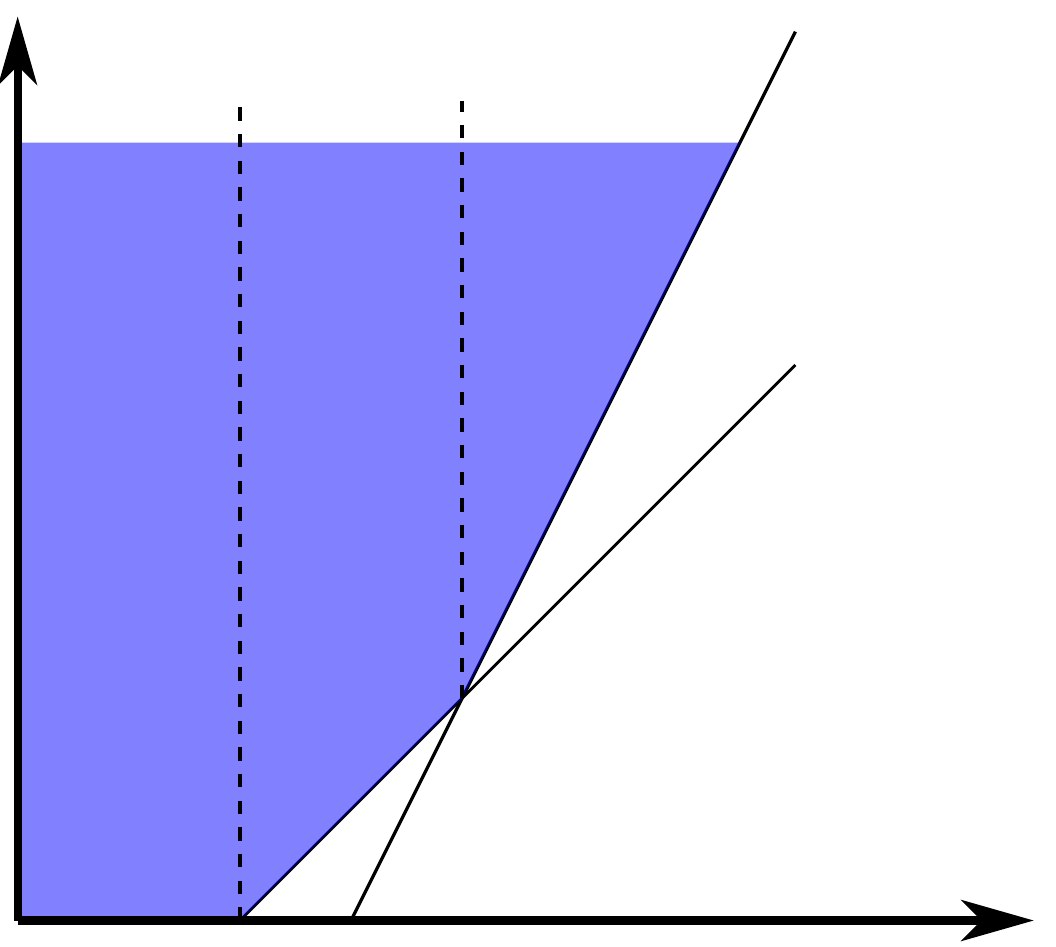} \qquad
\subfloat[][BCFW triangulation]{\label{5pt_ext}
\def\svgwidth{4,4cm}
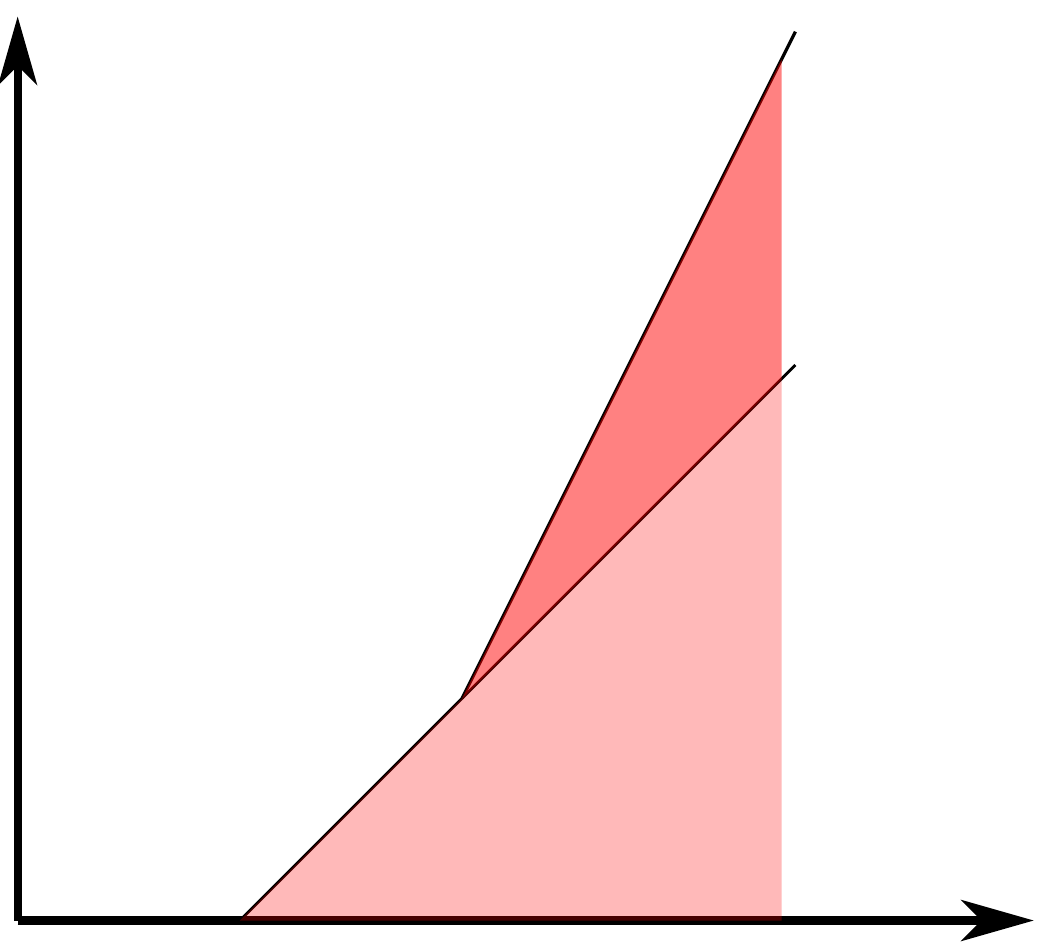}
\caption{The domain $\mathcal D^{(2)}_5$ and the two ways of constructing it.}\label{}
\end{figure}

When the number of points is increased, the presence of more $\theta$-functions guarantees that the domain of integration shrinks, as was already advocated in \cite{Arkani-Hamed:2014dca}. Let us show it on the five-point example. For concreteness, let us choose the following positive configuration
\begin{equation}
Z_4^A=(1,-1,1)\,,\qquad\qquad Z_5^A=(3,-2,1)\,,
\end{equation}
determining the integration domain in Fig.~\ref{Fig.m25points}.
As before, we can construct both an internal and an external triangulation (Fig.~\ref{5pt_int} and \ref{5pt_ext}, respectively), yielding the known result
\begin{equation}
\Omega^{(2)}_{5,1} = \{3\} + \{4\} + \{5\} = [123] + [134] + [145]\,.
\end{equation}

\bigskip

\begin{figure}[ht]
\centering
\def\svgwidth{5cm}
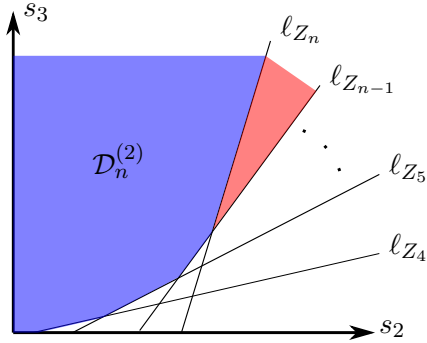
\caption{Generic domain of integration for $n$ points. We marked in red the wedge which evaluates to (minus) the R-invariant $[1\,n-1\,n]$.}\label{Fig.npoints}
\end{figure}

From these examples, we see an elegant pattern emerging. For $m=2$, the second summand in \eqref{intdomains} is in fact just an integral over the wedge
\begin{equation}\label{triangles}
\mathcal D^{(2)}_{n-1}\cap\{s\cdot Z_n  \leq 0\} = \{s\cdot Z_{n-1}  > 0\}\cap\{s\cdot Z_n \leq 0\}\,,
\end{equation}
depicted as the red area in Fig.~\ref{Fig.npoints}. It evaluates to the following R-invariant
\begin{equation}
 \int_{\{s\cdot Z_{n-1}  > 0\}\cap\{s\cdot Z_n \leq 0\}} ds \;(s\cdot Y)^{-3} = -[1\,n-1 \,n] \,.
\end{equation}
This gives a relation between the volume integral \eqref{volume3} and the BCFW decomposition of amplitudes for $k=1$. However, as we pointed out, there is no need to perform this triangulation in order to evaluate the integral \eqref{volume3}. This fact is even more relevant in the $m=4$ case, where the BCFW triangulation is more complicated. 


\subsection{Volume in the \texorpdfstring{$m=4$}{} case}

For any number of points $n$, the volume $\Omega_{n,1}^{(4)}$ is given by \cite{ArkaniHamed:2010gg}
\begin{equation}\label{answer4}
\Omega_{n,1}^{(4)}=\sum_{i<j}\, [1\,i\,i+1\,j\,j+1]\,,
\end{equation}
where the R-invariants are defined as
\begin{equation}
 [i\,j\,k\,l\,m]:=\frac{\langle i\,j\,k\,l\,m\rangle^4}{\langle Y \,i\,j\,k\,l\rangle \langle Y \,j\,k\,l\,m\rangle \langle Y \,k\,l\,m\,i\rangle \langle Y \,l\,m\,i\,j\rangle\langle Y \,m\,i\,j\,k\rangle}\,.
\end{equation}
In the following we will check that formula \eqref{volumem4} yields this result.

The simplest NMHV amplitude for $m=4$ is for five particles. This case is similar to the three-point volume for $m=2$, since there are no $\theta$-functions in the integrand of \eqref{volumem4}: 
\begin{equation}
\Omega^{(4)}_{5,1} = \int\displaylimits_0^{+\infty} ds_2 \int\displaylimits_0^{+\infty}ds_3 \int\displaylimits_0^{+\infty}ds_4\int\displaylimits_0^{+\infty} ds_5  \,\frac{4!}{ \left(Y^1 + s_2\,Y^2   +s_3\,Y^3  +s_4\, Y^4  +s_5\, Y^5  \right)^5}\,.
\end{equation}
The domain of integration is just the region of the four-dimensional real space where all coordinates are positive.
The usual argument ensures that the integrand is completely well defined, namely, it has no poles in $\mathcal D^{(4)}_5$. Computing the integral, we find
\begin{equation}
\Omega^{(4)}_{5,1}= \frac{1}{Y^1 Y^2 Y^3 Y^4 Y^5}\,,
\end{equation}
which lifts to the non-fixed form
\begin{equation}
\Omega^{(4)}_{5,1}=\frac{\langle 12345 \rangle^4}{\langle Y1234 \rangle \langle Y2345 \rangle\langle Y3451 \rangle\langle Y4512\rangle\langle Y5123\rangle} = [12345]\,,
\end{equation}
i.e. the correct result.

\begin{figure}[ht]
\centering
\subfloat[][R-invariant \mbox{[12356]}]{\label{R12356}
\def\svgwidth{4cm}
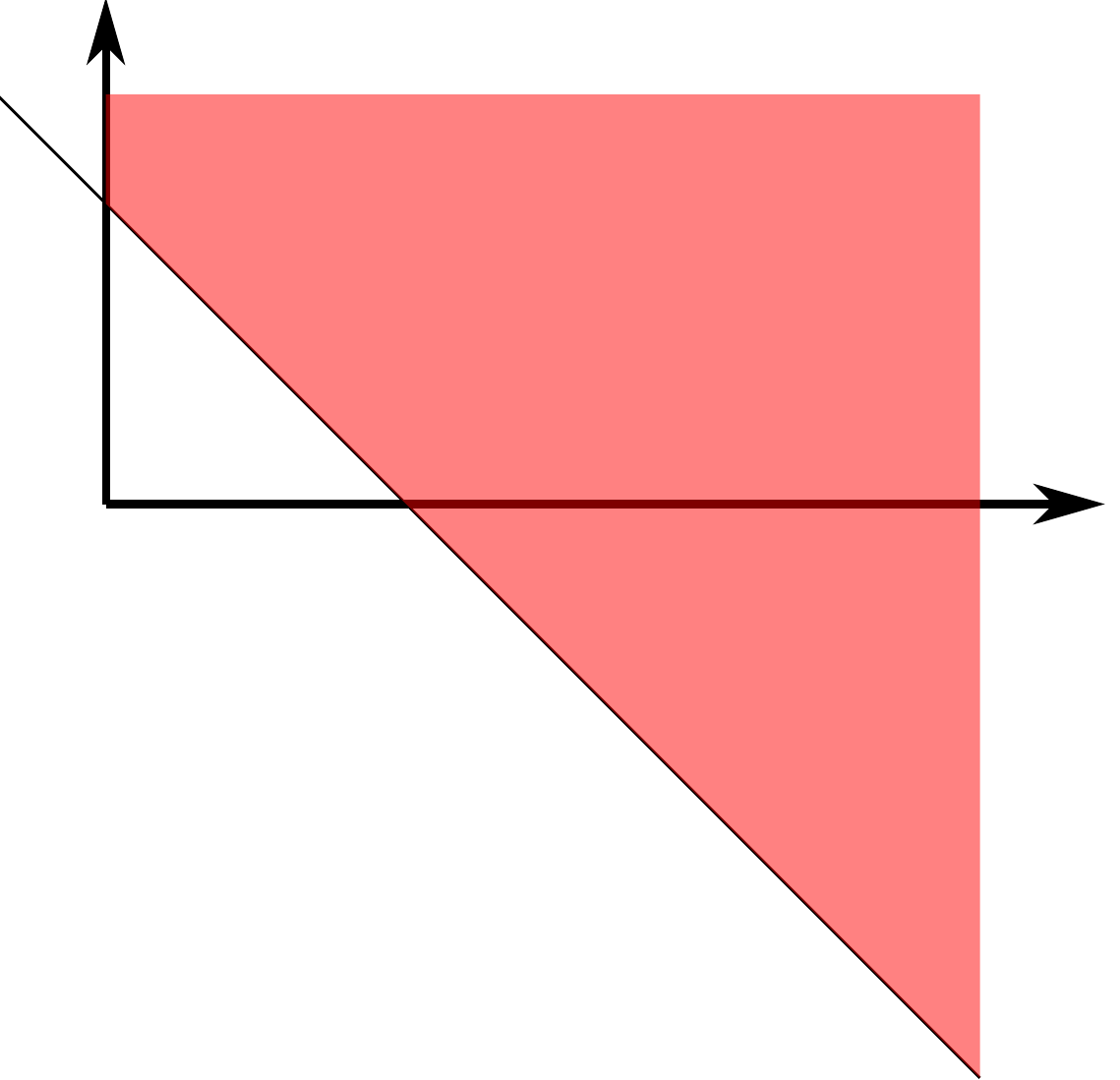} \qquad\qquad
\subfloat[][R-invariant \mbox{$[$13456$]$} ]{\label{R13456}
\def\svgwidth{4cm}
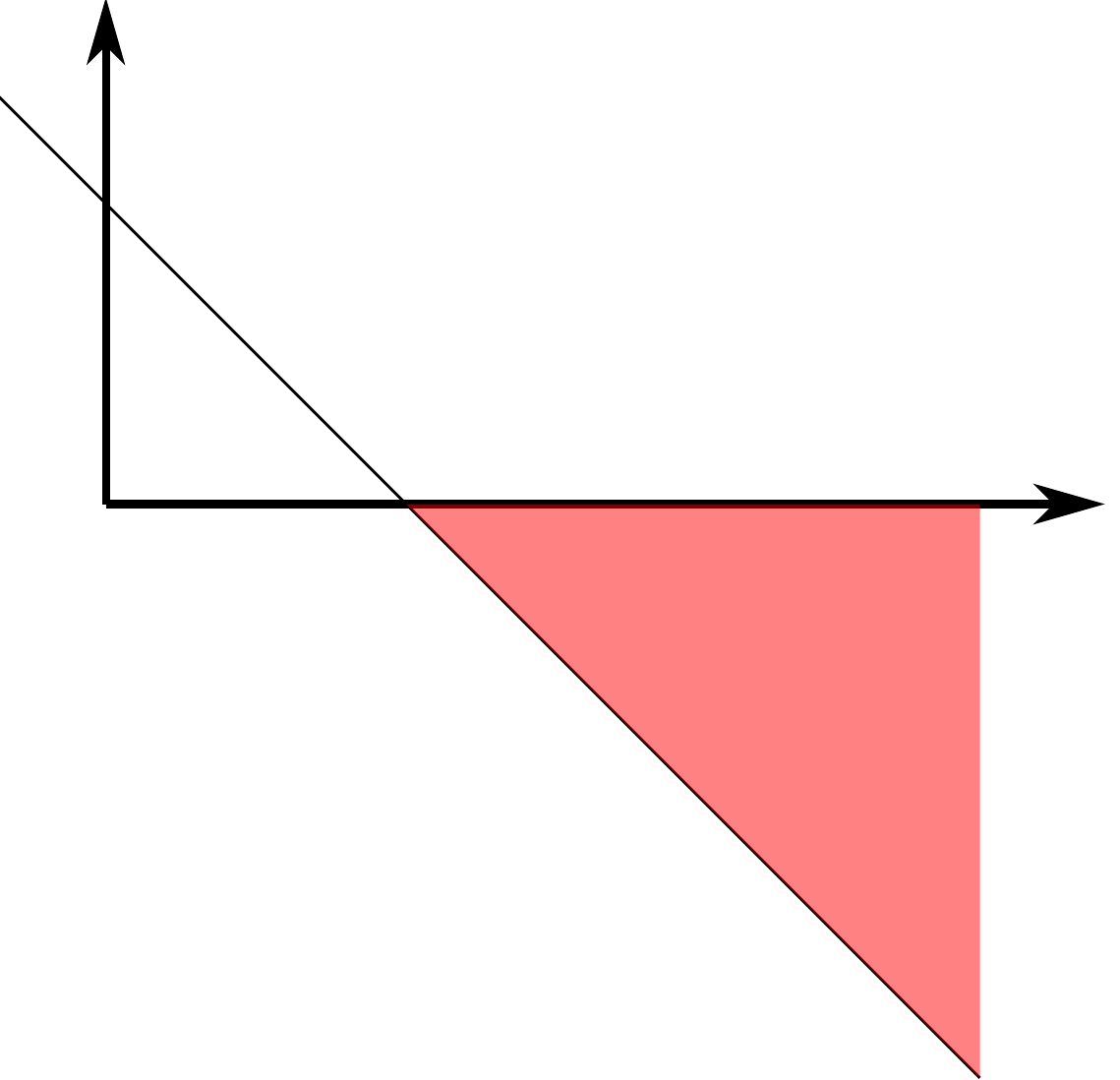}
\caption{Two contributions to the region of integration for $\Omega_{6,1}$.}\label{Fig.RinvDomains5pt}
\end{figure}

\begin{figure}[ht]
\centering
\def\svgwidth{\columnwidth}
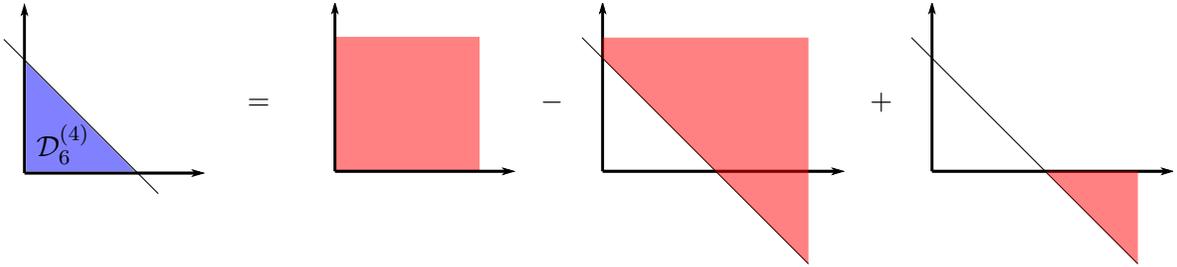
\caption{Domains of integration for six points and $m=4$}\label{Fig.m45points}
\end{figure}

For six points the formula \eqref{volumem4} reads
\begin{equation}\label{6pointsm4}
\Omega^{(4)}_{6,1} = 4! \int\displaylimits_0^{+\infty} ds_2 \int\displaylimits_0^{+\infty}ds_3 \int\displaylimits_0^{+\infty}ds_4\int\displaylimits_0^{+\infty} ds_5  \,\frac{\theta(Z_6^1 + s_2\,Z_6^2   +s_3\,Z_6^3  +s_4\, Z_6^4  +s_5\, Z_6^5)}{ \left(Y^1 + s_2\,Y^2   +s_3\,Y^3  +s_4\, Y^4  +s_5\, Y^5  \right)^5}\,.
\end{equation}
To simplify the discussion, we can again choose a particular positive configuration of external data. Let
\begin{equation}
Z_6^A=(1,-1,1,-1,1)\,,
\end{equation}
so that the $\theta$-function defines the hyperplane $\ell_{Z_6}: 1-s_2+s_3 -s_4 +s_5 = 0$. Solving the constraint, we can rewrite \eqref{6pointsm4} as
\begin{equation}\label{6pointsm4p}
\Omega^{(4)}_{6,1}= 4!\int\displaylimits_0^{+\infty} ds_3 \int\displaylimits_0^{+\infty} ds_5 \int\displaylimits_0^{1+s_3+s_5} ds_2 \int\displaylimits_0^{ 1-s_2+s_3+s_5} ds_4\,(s\cdot Y)^{-5}\,,
\end{equation}
which can be easily evaluated and agrees with the correct result for six-point NMHV amplitude \eqref{answer4}. In order to relate the integral \eqref{6pointsm4p} term-by-term with the BCFW recursion result
\begin{equation}
\Omega^{(4)}_{6,1}=[12345]+[12356]+[13456]\,,
\end{equation}
 let us observe that
\begin{align}
[12345] &=4! \int\displaylimits_0^{+\infty} ds_3\int\displaylimits_0^{+\infty} ds_5 \int\displaylimits_0^{+\infty} ds_2 \int\displaylimits_0^{+\infty} ds_4 \,(s\cdot Y)^{-5}\,, \\
[12356] &= - 4!\int\displaylimits_0^{+\infty} ds_3 \int\displaylimits_0^{+\infty} ds_5 \int\displaylimits_0^{+\infty} ds_2 \int\displaylimits_{1-s_2+s_3+s_5}^{+\infty} ds_4\,(s\cdot Y)^{-5} \,,\\
[13456] &= 4!\int\displaylimits_0^{+\infty} ds_3 \int\displaylimits_0^{+\infty} ds_5  \int\displaylimits_{1+s_3+s_5}^{+\infty} ds_2\int\displaylimits_{1-s_2+s_3+s_5}^{0} ds_4 \,(s\cdot Y)^{-5} \,.
\end{align}
It is enough to focus on the integration regions in the $(s_2,s_4)$ plane since the remaining two variables are integrated over $(0,+\infty)$ in all cases. Then, the domain of $[12345]$ is simply the positive quadrant, whereas those of $[12356]$ and $[13456]$ are depicted in Fig.~\ref{Fig.RinvDomains5pt}: here we solve the condition defining the hyperplane $\ell_{Z_6}$ as $s_4 = -s_2 + a$, where $a = 1+s_3+s_5$ is guaranteed to be positive. Fig.~\ref{Fig.m45points} shows that the various domains correctly add up to the integration region of $\Omega^{(4)}_{6,1}$, as in \eqref{6pointsm4p}.

For higher number of points the relation to BCFW recursion is more obscure since one has to study the full four-dimensional space in order to identify proper triangles. In particular, adding new particles does not simply correspond to removing a single triangle as in Fig.~\ref{Fig.npoints}, since \eqref{triangles} does not hold anymore. This can be traced back to the difference between formulas \eqref{answer2} and \eqref{answer4}: for $m=2$ we always add one R-invariant when increasing the number of particles by one, while for $m=4$ we need $n-4$ new contributions. However, thanks to formula \eqref{volumem4}, we can be cavalier about this, since the volume can be computed directly without any reference to triangulations.  
%


\subsection{First look at higher-helicity amplituhedron volumes}

Encouraged by the success of finding the volume formula for $k=1$,  we would like to see whether it is also possible to apply a similar approach to N$^k$MHV amplitudes for $k>1$. Solutions of several higher-order systems of Capelli equations supplemented by invariance and certain scaling properties can be found in the literature \cite{Oshima:1995capelliidentities}. Their integral representations in terms of dual Grassmannian variables are finite and can be computed for any value of the parameter $n$. Unfortunately, these cannot be interpreted as amplitudes due to scaling properties different from \eqref{scalingglobal}. However, let us proceed in a similar spirit as for $k=1$ and try to find what we can learn about the possible form of the solutions. 

First of all, similarly to the $k=1$ case, the Capelli equations introduce two sets of natural variables and a general form of the solution can be written as
\begin{equation}\label{Fouriergeneral}
\Omega^{(m)}_{n,k}(Y,Z)= \int ds_A^\alpha \,d\tilde s^i_\alpha\,\, e^{i\, s_A^\alpha\,Y^A_\alpha - i\, s_A^\alpha \,Z_i^A \,\tilde s^i_\alpha } \; F(s,\tilde s) \,.
\end{equation}
Then, demanding that all $Y_\alpha^A$ localize on hyperplanes defined by $Z_i^A$, namely
\begin{equation}
Y_{\alpha}^{A}=\tilde s^{i}_{\alpha} Z_{i}^{A}\,,
\end{equation}
we get again that the function $F(s,\tilde s)$ is independent of $s^\alpha_A$ and we can provide a representation purely in terms of dual Grassmannian coordinates:
\begin{equation}\label{omegakbigger}
\Omega^{(m)}_{n,k}(Y,Z)= \int ds_A^\alpha \,\, e^{i\, s_A^\alpha\,Y^A_\alpha }\,\tilde F(s^\alpha_A  \,Z^A_i)\,.
\end{equation}
The function $\tilde F(s^\alpha_A  \,Z^A_i)$ depends on $k\cdot n$ variables and the formula \eqref{scalingYdiff} implies that it depends on them through their $SL(k)$-invariant combinations 
\begin{equation}\label{brackets}
\{ i_1,\ldots ,i_k\}=\det ((s\cdot Z)_{i_1},\ldots ,(s\cdot Z)_{i_k})\,,
\end{equation}
where the compact notation $(s\cdot Z)_{i}$ is to be understood as in formula \eqref{omegakbigger}. Finally, we can use the scaling \eqref{scalingZdiff} and \eqref{scalingYdiff} to further restrict possible functions. It turns out, however, that this does not fix the final answer uniquely, since for $k>1$ we can form non-trivial cross-ratios from the brackets \eqref{brackets} and $\tilde F(s^\alpha_A  \,Z^A_i)$ could be in principle any function of these cross-ratios. 
In order to proceed further we should supplement known symmetries by an equivalent of the level-one Yangian invariance. However, at the moment the form of level-one generators is not known in the bosonized momentum twistor space.


\subsection{Deformed amplituhedron volume}

Finally, we would like to comment on possible natural deformations of the equations we studied so far, analogous to those introduced in the context of amplitudes in \cite{Ferro:2012xw,Ferro:2013dga}. For $k=1$ this amounts to more general scaling properties and the formula  \eqref{scalingZdiff} is replaced by
\begin{equation}\label{deformedscaling}
\sum_{A=1}^{m+1} Z_{i}^{A}\frac{\partial}{\partial Z_{i}^A}\,\Omega^{(m)}_{n,1}(Y,Z)=\alpha_i \,\Omega^{(m)}_{n,1}(Y,Z)\,,\qquad \text{for }i=1,\ldots,n\,,
\end{equation}
with 
\begin{equation}
\sum_{i=1}^n \alpha_i=0 \,.
\end{equation}
Let us remark that we only modify the weight of the variables $Z_i$ to match the deformed top-cell Grassmannian integral in \cite{Ferro:2014gca,Bargheer:2014mxa}.
In the context of scattering amplitudes the complex numbers $\alpha_i$ are related to the inhomogeneities of the integrable spin chain, as explained in \cite{Frassek:2013xza, Kanning:2014maa}. Indeed, the level-one Yangian generators \eqref{Yangiangen} get modified by local terms with inhomogeneities. In this generalized case, the solution to \eqref{invariancediff}, \eqref{Capelli} and \eqref{deformedscaling} can be also found in \cite{GelfandGraev} and reads
\begin{equation}\label{deformedampl}
\Omega^{(m)}_{n,1}(Y,Z)= \int ds_A \, e^{i \,s_A\,Y^A } \prod_i  (s_A \,Z^A_i)_+^{\alpha_i} \; ,
\end{equation}
where
\begin{equation}
x_+^\alpha=\begin{cases}x^\alpha,&x\geq 0\\0,&x<0\end{cases} \; .
\end{equation}
The integral \eqref{deformedampl} is a GKZ hypergeometric function and its properties were studied in e.g. \cite{GelfandGraev}. Importantly, it is convergent for $\alpha_i$ close to zero and can be evaluated explicitly. One notices that in the limit $\alpha_i\to 0$ the integral \eqref{deformedampl} smoothly approaches the integral \eqref{volume1}.


\section{Discussion and outlook}

In this paper we present a new approach to finding the volume of the tree amplituhedron based on its symmetries. This leads us to the novel formula \eqref{volume3} for the tree-level NMHV amplitudes allowing their evaluation without any reference to triangulations. In particular, the symmetries fix the volume formula to be an integral over the dual Grassmannian, which can be identified with the dual bosonized momentum twistor space. This suggests a natural generalization to higher-$k$ amplitudes,
leading to a natural framework where to write the volume as an integral over a dual Grassmannian.
In that case, however, we do not have enough symmetry to fix the final formula completely. This calls for further studies and in particular it raises the question on how to realize the Yangian symmetry directly in the bosonized momentum twistor space.


\section*{Acknowledgements}
We are grateful to Andrew Hodges and Nils Kanning for useful discussions. L.F. is supported by the Elitenetwork of Bavaria. T.L. is supported by ERC STG grant 306260.


\bibliographystyle{nb}
\bibliography{volume}

\end{document}

%% file: Images/NMHV_m2_3pt.pdf_tex
\begingroup%
  \makeatletter%
  \providecommand\color[2][]{%
    \errmessage{(Inkscape) Color is used for the text in Inkscape, but the package 'color.sty' is not loaded}%
    \renewcommand\color[2][]{}%
  }%
  \providecommand\transparent[1]{%
    \errmessage{(Inkscape) Transparency is used (non-zero) for the text in Inkscape, but the package 'transparent.sty' is not loaded}%
    \renewcommand\transparent[1]{}%
  }%
  \providecommand\rotatebox[2]{#2}%
  \ifx\svgwidth\undefined%
    \setlength{\unitlength}{304.10400391bp}%
    \ifx\svgscale\undefined%
      \relax%
    \else%
      \setlength{\unitlength}{\unitlength * \real{\svgscale}}%
    \fi%
  \else%
    \setlength{\unitlength}{\svgwidth}%
  \fi%
  \global\let\svgwidth\undefined%
  \global\let\svgscale\undefined%
  \makeatother%
  \begin{picture}(1,0.89658802)%
    \put(0,0){\includegraphics[width=\unitlength]{NMHV_m2_3pt.pdf}}%
    \put(0.99009275,0.02481694){\color[rgb]{0,0,0}\makebox(0,0)[lb]{\smash{$s_2$}}}%
    \put(0.04304833,0.86663421){\color[rgb]{0,0,0}\makebox(0,0)[lb]{\smash{$s_3$}}}%
    \put(0.38503659,0.36680505){\color[rgb]{0,0,0}\makebox(0,0)[lb]{\smash{$\mathcal D^{(2)}_3$}}}%
  \end{picture}%
\endgroup%

%% file: Images/NMHV_m2_4pt.pdf_tex
\begingroup%
  \makeatletter%
  \providecommand\color[2][]{%
    \errmessage{(Inkscape) Color is used for the text in Inkscape, but the package 'color.sty' is not loaded}%
    \renewcommand\color[2][]{}%
  }%
  \providecommand\transparent[1]{%
    \errmessage{(Inkscape) Transparency is used (non-zero) for the text in Inkscape, but the package 'transparent.sty' is not loaded}%
    \renewcommand\transparent[1]{}%
  }%
  \providecommand\rotatebox[2]{#2}%
  \ifx\svgwidth\undefined%
    \setlength{\unitlength}{304.10400391bp}%
    \ifx\svgscale\undefined%
      \relax%
    \else%
      \setlength{\unitlength}{\unitlength * \real{\svgscale}}%
    \fi%
  \else%
    \setlength{\unitlength}{\svgwidth}%
  \fi%
  \global\let\svgwidth\undefined%
  \global\let\svgscale\undefined%
  \makeatother%
  \begin{picture}(1,0.89658802)%
    \put(0,0){\includegraphics[width=\unitlength]{NMHV_m2_4pt.pdf}}%
    \put(0.99009275,0.02481694){\color[rgb]{0,0,0}\makebox(0,0)[lb]{\smash{$s_2$}}}%
    \put(0.04304833,0.86663421){\color[rgb]{0,0,0}\makebox(0,0)[lb]{\smash{$s_3$}}}%
    \put(0.25350264,0.4720322){\color[rgb]{0,0,0}\makebox(0,0)[lb]{\smash{$\mathcal D^{(2)}_4$}}}%
    \put(0.99009275,0.76140685){\color[rgb]{0,0,0}\makebox(0,0)[lb]{\smash{$\ell_{Z_4}$}}}%
  \end{picture}%
\endgroup%

%% file: Images/NMHV_m2_4pt_int.pdf_tex
\begingroup%
  \makeatletter%
  \providecommand\color[2][]{%
    \errmessage{(Inkscape) Color is used for the text in Inkscape, but the package 'color.sty' is not loaded}%
    \renewcommand\color[2][]{}%
  }%
  \providecommand\transparent[1]{%
    \errmessage{(Inkscape) Transparency is used (non-zero) for the text in Inkscape, but the package 'transparent.sty' is not loaded}%
    \renewcommand\transparent[1]{}%
  }%
  \providecommand\rotatebox[2]{#2}%
  \ifx\svgwidth\undefined%
    \setlength{\unitlength}{304.10400391bp}%
    \ifx\svgscale\undefined%
      \relax%
    \else%
      \setlength{\unitlength}{\unitlength * \real{\svgscale}}%
    \fi%
  \else%
    \setlength{\unitlength}{\svgwidth}%
  \fi%
  \global\let\svgwidth\undefined%
  \global\let\svgscale\undefined%
  \makeatother%
  \begin{picture}(1,0.89658802)%
    \put(0,0){\includegraphics[width=\unitlength]{NMHV_m2_4pt_int.pdf}}%
    \put(0.99009275,0.02481694){\color[rgb]{0,0,0}\makebox(0,0)[lb]{\smash{$s_2$}}}%
    \put(0.04304833,0.86663421){\color[rgb]{0,0,0}\makebox(0,0)[lb]{\smash{$s_3$}}}%
    \put(0.97693935,0.78771364){\color[rgb]{0,0,0}\makebox(0,0)[lb]{\smash{$\ell_{Z_4}$}}}%
    \put(0.06935511,0.51149239){\color[rgb]{0,0,0}\makebox(0,0)[lb]{\smash{$\{3\}$}}}%
    \put(0.39818998,0.51149239){\color[rgb]{0,0,0}\makebox(0,0)[lb]{\smash{$\{4\}$}}}%
  \end{picture}%
\endgroup%

%% file: Images/NMHV_m2_4pt_ext.pdf_tex
\begingroup%
  \makeatletter%
  \providecommand\color[2][]{%
    \errmessage{(Inkscape) Color is used for the text in Inkscape, but the package 'color.sty' is not loaded}%
    \renewcommand\color[2][]{}%
  }%
  \providecommand\transparent[1]{%
    \errmessage{(Inkscape) Transparency is used (non-zero) for the text in Inkscape, but the package 'transparent.sty' is not loaded}%
    \renewcommand\transparent[1]{}%
  }%
  \providecommand\rotatebox[2]{#2}%
  \ifx\svgwidth\undefined%
    \setlength{\unitlength}{304.10400391bp}%
    \ifx\svgscale\undefined%
      \relax%
    \else%
      \setlength{\unitlength}{\unitlength * \real{\svgscale}}%
    \fi%
  \else%
    \setlength{\unitlength}{\svgwidth}%
  \fi%
  \global\let\svgwidth\undefined%
  \global\let\svgscale\undefined%
  \makeatother%
  \begin{picture}(1,0.89658802)%
    \put(0,0){\includegraphics[width=\unitlength]{NMHV_m2_4pt_ext.pdf}}%
    \put(0.99009275,0.02481694){\color[rgb]{0,0,0}\makebox(0,0)[lb]{\smash{$s_2$}}}%
    \put(0.04304833,0.86663421){\color[rgb]{0,0,0}\makebox(0,0)[lb]{\smash{$s_3$}}}%
    \put(0.5954909,0.18265752){\color[rgb]{0,0,0}\makebox(0,0)[lb]{\smash{$-[134]$}}}%
    \put(0.97693935,0.78771364){\color[rgb]{0,0,0}\makebox(0,0)[lb]{\smash{$\ell_{Z_4}$}}}%
  \end{picture}%
\endgroup%

%% file: Images/NMHV_m2_5pt.pdf_tex
\begingroup%
  \makeatletter%
  \providecommand\color[2][]{%
    \errmessage{(Inkscape) Color is used for the text in Inkscape, but the package 'color.sty' is not loaded}%
    \renewcommand\color[2][]{}%
  }%
  \providecommand\transparent[1]{%
    \errmessage{(Inkscape) Transparency is used (non-zero) for the text in Inkscape, but the package 'transparent.sty' is not loaded}%
    \renewcommand\transparent[1]{}%
  }%
  \providecommand\rotatebox[2]{#2}%
  \ifx\svgwidth\undefined%
    \setlength{\unitlength}{304.10400391bp}%
    \ifx\svgscale\undefined%
      \relax%
    \else%
      \setlength{\unitlength}{\unitlength * \real{\svgscale}}%
    \fi%
  \else%
    \setlength{\unitlength}{\svgwidth}%
  \fi%
  \global\let\svgwidth\undefined%
  \global\let\svgscale\undefined%
  \makeatother%
  \begin{picture}(1,0.89658802)%
    \put(0,0){\includegraphics[width=\unitlength]{NMHV_m2_5pt.pdf}}%
    \put(0.99009275,0.02481694){\color[rgb]{0,0,0}\makebox(0,0)[lb]{\smash{$s_2$}}}%
    \put(0.04304833,0.86663421){\color[rgb]{0,0,0}\makebox(0,0)[lb]{\smash{$s_3$}}}%
    \put(0.22719585,0.57725936){\color[rgb]{0,0,0}\makebox(0,0)[lb]{\smash{$\mathcal D^{(2)}_5$}}}%
    \put(0.77963843,0.84032722){\color[rgb]{0,0,0}\makebox(0,0)[lb]{\smash{$\ell_{Z_5}$}}}%
    \put(0.83225201,0.57725936){\color[rgb]{0,0,0}\makebox(0,0)[lb]{\smash{$\ell_{Z_4}$}}}%
  \end{picture}%
\endgroup%

%% file: Images/NMHV_m2_5pt_int.pdf_tex
\begingroup%
  \makeatletter%
  \providecommand\color[2][]{%
    \errmessage{(Inkscape) Color is used for the text in Inkscape, but the package 'color.sty' is not loaded}%
    \renewcommand\color[2][]{}%
  }%
  \providecommand\transparent[1]{%
    \errmessage{(Inkscape) Transparency is used (non-zero) for the text in Inkscape, but the package 'transparent.sty' is not loaded}%
    \renewcommand\transparent[1]{}%
  }%
  \providecommand\rotatebox[2]{#2}%
  \ifx\svgwidth\undefined%
    \setlength{\unitlength}{304.10400391bp}%
    \ifx\svgscale\undefined%
      \relax%
    \else%
      \setlength{\unitlength}{\unitlength * \real{\svgscale}}%
    \fi%
  \else%
    \setlength{\unitlength}{\svgwidth}%
  \fi%
  \global\let\svgwidth\undefined%
  \global\let\svgscale\undefined%
  \makeatother%
  \begin{picture}(1,0.89658802)%
    \put(0,0){\includegraphics[width=\unitlength]{NMHV_m2_5pt_int.pdf}}%
    \put(0.99009275,0.02481694){\color[rgb]{0,0,0}\makebox(0,0)[lb]{\smash{$s_2$}}}%
    \put(0.04304833,0.86663421){\color[rgb]{0,0,0}\makebox(0,0)[lb]{\smash{$s_3$}}}%
    \put(0.77963843,0.52464578){\color[rgb]{0,0,0}\makebox(0,0)[lb]{\smash{$\ell_{Z_4}$}}}%
    \put(0.77963843,0.84032722){\color[rgb]{0,0,0}\makebox(0,0)[lb]{\smash{$\ell_{Z_5}$}}}%
    \put(0.26665604,0.65617973){\color[rgb]{0,0,0}\makebox(0,0)[lb]{\smash{$\{4\}$}}}%
    \put(0.05620172,0.65617973){\color[rgb]{0,0,0}\makebox(0,0)[lb]{\smash{$\{3\}$}}}%
    \put(0.49026375,0.65617973){\color[rgb]{0,0,0}\makebox(0,0)[lb]{\smash{$\{5\}$}}}%
  \end{picture}%
\endgroup%

%% file: Images/NMHV_m2_5pt_ext.pdf_tex
\begingroup%
  \makeatletter%
  \providecommand\color[2][]{%
    \errmessage{(Inkscape) Color is used for the text in Inkscape, but the package 'color.sty' is not loaded}%
    \renewcommand\color[2][]{}%
  }%
  \providecommand\transparent[1]{%
    \errmessage{(Inkscape) Transparency is used (non-zero) for the text in Inkscape, but the package 'transparent.sty' is not loaded}%
    \renewcommand\transparent[1]{}%
  }%
  \providecommand\rotatebox[2]{#2}%
  \ifx\svgwidth\undefined%
    \setlength{\unitlength}{304.10400391bp}%
    \ifx\svgscale\undefined%
      \relax%
    \else%
      \setlength{\unitlength}{\unitlength * \real{\svgscale}}%
    \fi%
  \else%
    \setlength{\unitlength}{\svgwidth}%
  \fi%
  \global\let\svgwidth\undefined%
  \global\let\svgscale\undefined%
  \makeatother%
  \begin{picture}(1,0.89658802)%
    \put(0,0){\includegraphics[width=\unitlength]{NMHV_m2_5pt_ext.pdf}}%
    \put(0.99009275,0.02481694){\color[rgb]{0,0,0}\makebox(0,0)[lb]{\smash{$s_2$}}}%
    \put(0.04304833,0.86663421){\color[rgb]{0,0,0}\makebox(0,0)[lb]{\smash{$s_3$}}}%
    \put(0.67441127,0.64302633){\color[rgb]{0,0,0}\makebox(0,0)[lb]{\smash{$-[145]$}}}%
    \put(0.46395696,0.15635073){\color[rgb]{0,0,0}\makebox(0,0)[lb]{\smash{$-[134]$}}}%
    \put(0.77963843,0.52464578){\color[rgb]{0,0,0}\makebox(0,0)[lb]{\smash{$\ell_{Z_4}$}}}%
    \put(0.77963843,0.84032722){\color[rgb]{0,0,0}\makebox(0,0)[lb]{\smash{$\ell_{Z_5}$}}}%
  \end{picture}%
\endgroup%

%% file: Images/NMHV_m2_generic.pdf_tex
\begingroup%
  \makeatletter%
  \providecommand\color[2][]{%
    \errmessage{(Inkscape) Color is used for the text in Inkscape, but the package 'color.sty' is not loaded}%
    \renewcommand\color[2][]{}%
  }%
  \providecommand\transparent[1]{%
    \errmessage{(Inkscape) Transparency is used (non-zero) for the text in Inkscape, but the package 'transparent.sty' is not loaded}%
    \renewcommand\transparent[1]{}%
  }%
  \providecommand\rotatebox[2]{#2}%
  \ifx\svgwidth\undefined%
    \setlength{\unitlength}{304.10400391bp}%
    \ifx\svgscale\undefined%
      \relax%
    \else%
      \setlength{\unitlength}{\unitlength * \real{\svgscale}}%
    \fi%
  \else%
    \setlength{\unitlength}{\svgwidth}%
  \fi%
  \global\let\svgwidth\undefined%
  \global\let\svgscale\undefined%
  \makeatother%
  \begin{picture}(1,0.89658802)%
    \put(0,0){\includegraphics[width=\unitlength]{NMHV_m2_generic.pdf}}%
    \put(0.99009275,0.02481694){\color[rgb]{0,0,0}\makebox(0,0)[lb]{\smash{$s_2$}}}%
    \put(0.04304833,0.86663421){\color[rgb]{0,0,0}\makebox(0,0)[lb]{\smash{$s_3$}}}%
    \put(1.01639953,0.2352711){\color[rgb]{0,0,0}\makebox(0,0)[lb]{\smash{$\ell_{Z_4}$}}}%
    \put(1.01639953,0.44572541){\color[rgb]{0,0,0}\makebox(0,0)[lb]{\smash{$\ell_{Z_5}$}}}%
    \put(0.8585588,0.68248652){\color[rgb]{0,0,0}\makebox(0,0)[lb]{\smash{$\ell_{Z_{n-1}}$}}}%
    \put(0.72702485,0.81402043){\color[rgb]{0,0,0}\makebox(0,0)[lb]{\smash{$\ell_{Z_n}$}}}%
    \put(0.22719585,0.44572541){\color[rgb]{0,0,0}\makebox(0,0)[lb]{\smash{$\mathcal D^{(2)}_n$}}}%
  \end{picture}%
\endgroup%

%% file: Images/NMHV_m4_5ptR12356.pdf_tex
\begingroup%
  \makeatletter%
  \providecommand\color[2][]{%
    \errmessage{(Inkscape) Color is used for the text in Inkscape, but the package 'color.sty' is not loaded}%
    \renewcommand\color[2][]{}%
  }%
  \providecommand\transparent[1]{%
    \errmessage{(Inkscape) Transparency is used (non-zero) for the text in Inkscape, but the package 'transparent.sty' is not loaded}%
    \renewcommand\transparent[1]{}%
  }%
  \providecommand\rotatebox[2]{#2}%
  \ifx\svgwidth\undefined%
    \setlength{\unitlength}{328bp}%
    \ifx\svgscale\undefined%
      \relax%
    \else%
      \setlength{\unitlength}{\unitlength * \real{\svgscale}}%
    \fi%
  \else%
    \setlength{\unitlength}{\svgwidth}%
  \fi%
  \global\let\svgwidth\undefined%
  \global\let\svgscale\undefined%
  \makeatother%
  \begin{picture}(1,0.96341463)%
    \put(0,0){\includegraphics[width=\unitlength]{NMHV_m4_5ptR12356.pdf}}%
    \put(0.99696312,0.51341482){\color[rgb]{0,0,0}\makebox(0,0)[lb]{\smash{$s_2$}}}%
    \put(0.11891434,0.95243921){\color[rgb]{0,0,0}\makebox(0,0)[lb]{\smash{$s_4$}}}%
    \put(0.31403629,0.45243906){\color[rgb]{0,0,0}\makebox(0,0)[lb]{\smash{$a$}}}%
    \put(0.03354849,0.73292686){\color[rgb]{0,0,0}\makebox(0,0)[lb]{\smash{$a$}}}%
    \put(0.35062166,0.68414637){\color[rgb]{0,0,0}\makebox(0,0)[lb]{\smash{$-[12356]$}}}%
    \put(0.89940215,0.02560979){\color[rgb]{0,0,0}\makebox(0,0)[lb]{\smash{$\ell_{Z_6}$}}}%
  \end{picture}%
\endgroup%

%% file: Images/NMHV_m4_5ptR13456.pdf_tex
\begingroup%
  \makeatletter%
  \providecommand\color[2][]{%
    \errmessage{(Inkscape) Color is used for the text in Inkscape, but the package 'color.sty' is not loaded}%
    \renewcommand\color[2][]{}%
  }%
  \providecommand\transparent[1]{%
    \errmessage{(Inkscape) Transparency is used (non-zero) for the text in Inkscape, but the package 'transparent.sty' is not loaded}%
    \renewcommand\transparent[1]{}%
  }%
  \providecommand\rotatebox[2]{#2}%
  \ifx\svgwidth\undefined%
    \setlength{\unitlength}{328bp}%
    \ifx\svgscale\undefined%
      \relax%
    \else%
      \setlength{\unitlength}{\unitlength * \real{\svgscale}}%
    \fi%
  \else%
    \setlength{\unitlength}{\svgwidth}%
  \fi%
  \global\let\svgwidth\undefined%
  \global\let\svgscale\undefined%
  \makeatother%
  \begin{picture}(1,0.96341463)%
    \put(0,0){\includegraphics[width=\unitlength]{NMHV_m4_5ptR13456.pdf}}%
    \put(0.99696312,0.51341482){\color[rgb]{0,0,0}\makebox(0,0)[lb]{\smash{$s_2$}}}%
    \put(0.11891434,0.95243921){\color[rgb]{0,0,0}\makebox(0,0)[lb]{\smash{$s_4$}}}%
    \put(0.31403629,0.45243906){\color[rgb]{0,0,0}\makebox(0,0)[lb]{\smash{$a$}}}%
    \put(0.03354849,0.73292686){\color[rgb]{0,0,0}\makebox(0,0)[lb]{\smash{$a$}}}%
    \put(0.55793873,0.3670732){\color[rgb]{0,0,0}\makebox(0,0)[lb]{\smash{$[13456]$}}}%
    \put(0.89940215,-0.0231707){\color[rgb]{0,0,0}\makebox(0,0)[lb]{\smash{$\ell_{Z_6}$}}}%
  \end{picture}%
\endgroup%

%% file: Images/NMHV_m4_5pteq.pdf_tex
\begingroup%
  \makeatletter%
  \providecommand\color[2][]{%
    \errmessage{(Inkscape) Color is used for the text in Inkscape, but the package 'color.sty' is not loaded}%
    \renewcommand\color[2][]{}%
  }%
  \providecommand\transparent[1]{%
    \errmessage{(Inkscape) Transparency is used (non-zero) for the text in Inkscape, but the package 'transparent.sty' is not loaded}%
    \renewcommand\transparent[1]{}%
  }%
  \providecommand\rotatebox[2]{#2}%
  \ifx\svgwidth\undefined%
    \setlength{\unitlength}{908.6875bp}%
    \ifx\svgscale\undefined%
      \relax%
    \else%
      \setlength{\unitlength}{\unitlength * \real{\svgscale}}%
    \fi%
  \else%
    \setlength{\unitlength}{\svgwidth}%
  \fi%
  \global\let\svgwidth\undefined%
  \global\let\svgscale\undefined%
  \makeatother%
  \begin{picture}(1,0.22365059)%
    \put(0,0){\includegraphics[width=\unitlength]{NMHV_m4_5pteq.pdf}}%
    \put(0.2087489,0.1324988){\color[rgb]{0,0,0}\makebox(0,0)[lb]{\smash{$=$}}}%
    \put(0.45966024,0.1324988){\color[rgb]{0,0,0}\makebox(0,0)[lb]{\smash{$-$}}}%
    \put(0.74138525,0.1324988){\color[rgb]{0,0,0}\makebox(0,0)[lb]{\smash{$+$}}}%
    \put(0.02826881,0.09288122){\color[rgb]{0,0,0}\makebox(0,0)[lb]{\smash{$\mathcal D^{(4)}_6$}}}%
  \end{picture}%
\endgroup%

%% file: TowardsVolume_v2.bbl
\begin{thebibliography}{10}
\ifx\href\asklfhas\newcommand{\href}[2]{#2}\fi
\ifx\arxivref\asklfhas\newcommand{\arxivref}[2]{\href{http://arxiv.org/abs/#1}{#2}}\fi
\ifx\doiref\asklfhas\newcommand{\doiref}[2]{\href{http://dx.doi.org/#1}{#2}}\fi
\raggedright
\small
\parskip 0pt

\bibitem{Maldacena:1997re}
J.~M.~Maldacena,
\textit{``{The Large N limit of superconformal field theories and
  supergravity}''},
\textsf{\doiref{10.1023/A:1026654312961}{Int.~J.~Theor.~Phys.~38,~1113~(1999)}},
\texttt{\arxivref{hep-th/9711200}{hep-th/9711200}},
[Adv. Theor. Math. Phys.2,231(1998)].

\bibitem{Witten:2003nn}
E.~Witten,
\textit{``{Perturbative gauge theory as a string theory in twistor space}''},
\textsf{\doiref{10.1007/s00220-004-1187-3}{Commun.~Math.~Phys.~252,~189~(2004)}},
\texttt{\arxivref{hep-th/0312171}{hep-th/0312171}}.

\bibitem{ArkaniHamed:2010gg}
N.~Arkani-Hamed, J.~L.~Bourjaily, F.~Cachazo, A.~Hodges and J.~Trnka,
\textit{``{A Note on Polytopes for Scattering Amplitudes}''},
\textsf{\doiref{10.1007/JHEP04(2012)081}{JHEP~1204,~081~(2012)}},
\texttt{\arxivref{1012.6030}{arxiv:1012.6030}}.

\bibitem{ArkaniHamed:2012nw}
N.~Arkani-Hamed, J.~L.~Bourjaily, F.~Cachazo, A.~B.~Goncharov, A.~Postnikov and
  J.~Trnka,
\textit{``{Scattering Amplitudes and the Positive Grassmannian}''},
\texttt{\arxivref{1212.5605}{arxiv:1212.5605}}.

\bibitem{Arkani-Hamed:2013jha}
N.~Arkani-Hamed and J.~Trnka,
\textit{``{The Amplituhedron}''},
\textsf{\doiref{10.1007/JHEP10(2014)030}{JHEP~1410,~030~(2014)}},
\texttt{\arxivref{1312.2007}{arxiv:1312.2007}}.

\bibitem{Hodges:2009hk}
A.~Hodges,
\textit{``{Eliminating spurious poles from gauge-theoretic amplitudes}''},
\textsf{\doiref{10.1007/JHEP05(2013)135}{JHEP~1305,~135~(2013)}},
\texttt{\arxivref{0905.1473}{arxiv:0905.1473}}.

\bibitem{Drummond:2008vq}
J.~M.~Drummond, J.~Henn, G.~P.~Korchemsky and E.~Sokatchev,
\textit{``{Dual superconformal symmetry of scattering amplitudes in
  {$\mathcal{N}=\mathord{}$4} super-Yang--Mills theory}''},
\textsf{\doiref{10.1016/j.nuclphysb.2009.11.022}{Nucl.~Phys.~B828,~317~(2010)}},
\texttt{\arxivref{0807.1095}{arxiv:0807.1095}}.

\bibitem{Drummond:2009fd}
J.~M.~Drummond, J.~M.~Henn and J.~Plefka,
\textit{``{Yangian symmetry of scattering amplitudes in N=4 super Yang-Mills
  theory}''},
\textsf{\doiref{10.1088/1126-6708/2009/05/046}{JHEP~0905,~046~(2009)}},
\texttt{\arxivref{0902.2987}{arxiv:0902.2987}}.

\bibitem{ArkaniHamed:2008gz}
N.~Arkani-Hamed, F.~Cachazo and J.~Kaplan,
\textit{``{What is the Simplest Quantum Field Theory?}''},
\textsf{\doiref{10.1007/JHEP09(2010)016}{JHEP~1009,~016~(2010)}},
\texttt{\arxivref{0808.1446}{arxiv:0808.1446}}.

\bibitem{ArkaniHamed:2009vw}
N.~Arkani-Hamed, F.~Cachazo and C.~Cheung,
\textit{``{The Grassmannian Origin Of Dual Superconformal Invariance}''},
\textsf{\doiref{10.1007/JHEP03(2010)036}{JHEP~1003,~036~(2010)}},
\texttt{\arxivref{0909.0483}{arxiv:0909.0483}}.

\bibitem{Drummond:2010qh}
J.~Drummond and L.~Ferro,
\textit{``{Yangians, Grassmannians and T-duality}''},
\textsf{\doiref{10.1007/JHEP07(2010)027}{JHEP~1007,~027~(2010)}},
\texttt{\arxivref{1001.3348}{arxiv:1001.3348}}.

\bibitem{Drummond:2010uq}
J.~M.~Drummond and L.~Ferro,
\textit{``{The Yangian origin of the Grassmannian integral}''},
\textsf{\doiref{10.1007/JHEP12(2010)010}{JHEP~1012,~010~(2010)}},
\texttt{\arxivref{1002.4622}{arxiv:1002.4622}}.

\bibitem{ArkaniHamed:2009dn}
N.~Arkani-Hamed, F.~Cachazo, C.~Cheung and J.~Kaplan,
\textit{``{A Duality For The S Matrix}''},
\textsf{\doiref{10.1007/JHEP03(2010)020}{JHEP~1003,~020~(2010)}},
\texttt{\arxivref{0907.5418}{arxiv:0907.5418}}.

\bibitem{Mason:2009qx}
L.~Mason and D.~Skinner,
\textit{``{Dual Superconformal Invariance, Momentum Twistors and
  Grassmannians}''},
\textsf{\doiref{10.1088/1126-6708/2009/11/045}{JHEP~0911,~045~(2009)}},
\texttt{\arxivref{0909.0250}{arxiv:0909.0250}}.

\bibitem{Arkani-Hamed:2014dca}
N.~Arkani-Hamed, A.~Hodges and J.~Trnka,
\textit{``{Positive Amplitudes In The Amplituhedron}''},
\textsf{\doiref{10.1007/JHEP08(2015)030}{JHEP~1508,~030~(2015)}},
\texttt{\arxivref{1412.8478}{arxiv:1412.8478}}.

\bibitem{Korchemsky:2010ut}
G.~Korchemsky and E.~Sokatchev,
\textit{``{Superconformal invariants for scattering amplitudes in N=4 SYM
  theory}''},
\textsf{\doiref{10.1016/j.nuclphysb.2010.05.022}{Nucl.Phys.~B839,~377~(2010)}},
\texttt{\arxivref{1002.4625}{arxiv:1002.4625}}.

\bibitem{Ferro:2012xw}
L.~Ferro, T.~{\L}ukowski, C.~Meneghelli, J.~Plefka and M.~Staudacher,
\textit{``{Harmonic R-matrices for Scattering Amplitudes and Spectral
  Regularization}''},
\textsf{\doiref{10.1103/PhysRevLett.110.121602}{Phys.Rev.Lett.~110,~121602~(2013)}},
\texttt{\arxivref{1212.0850}{arxiv:1212.0850}}.

\bibitem{Ferro:2013dga}
L.~Ferro, T.~{\L}ukowski, C.~Meneghelli, J.~Plefka and M.~Staudacher,
\textit{``{Spectral Parameters for Scattering Amplitudes in N=4 Super
  Yang-Mills Theory}''},
\textsf{\doiref{10.1007/JHEP01(2014)094}{JHEP~1401,~094~(2014)}},
\texttt{\arxivref{1308.3494}{arxiv:1308.3494}}.

\bibitem{MR841131}
I.~Gelfand,
\textit{``General theory of hypergeometric functions''},
\textsf{Dokl.~Akad.~Nauk~SSSR~288,~14~(1986)}.

\bibitem{Aomoto:1975}
K.~Aomoto,
\textit{``Les \'equations aux diff\'erences lin\'eaires et les int\'egrales des
  fonctions multiformes''},
\textsf{J.~Fac.~Sci.~Univ.~Tokyo,~Sect.~IA~Math.~22,~271~(1975)}.

\bibitem{Aomoto}
M.~Kita and K.~Aomoto,
\textit{``Theory of hypergeometric functions''},
Springer-Verlag (2011).

\bibitem{Ferro:2014gca}
L.~Ferro, T.~{\L}ukowski and M.~Staudacher,
\textit{``{$\mathcal N=4$ scattering amplitudes and the deformed
  Grassmannian}''},
\textsf{\doiref{10.1016/j.nuclphysb.2014.10.012}{Nucl.~Phys.~B889,~192~(2014)}},
\texttt{\arxivref{1407.6736}{arxiv:1407.6736}}.

\bibitem{GelfandGraev}
I.~M.~Gelfand, M.~I.~Graev and V.~S.~Retakh,
\textit{``{General hypergeometric systems of equations and series of
  hypergeometric type}''},
\textsf{Uspekhi~Mat.~Nauk~SSSR~47:4,~3~(1992)}.

\bibitem{Oshima:1995capelliidentities}
T.~Oshima,
\textit{``Capelli Identities, Degenerate Series and Hypergeometric
  Functions''},
in: \textit{``Proceedings of a symposium on Representation Theory at Okinawa
  (1995)''},
{1-19}p.

\bibitem{gelʹfand1964generalized}
I.~Gelʹfand and G.~Shilov,
\textit{``Generalized Functions: Properties and operations, by I. M. Gelʹfand
  and G. E. Shilov, translated by E. Saletan''},
Academic Press (1964).

\bibitem{Nima:2014}
N.~Arkani-Hamed,
\textit{``The amplituhedron, scattering amplitudes and $\Psi_U$''},
New geometric structures in scattering amplitudes,
\href{http://people.maths.ox.ac.uk/lmason/NGSA14/Films/Nima-Arkani-Hamed.mp4}{\texttt{http://people.maths.ox.ac.uk/lmason/NGSA14/Films/Nima-Arkani-Hamed.mp4}}.

\bibitem{Bargheer:2014mxa}
T.~Bargheer, Y.-t.~Huang, F.~Loebbert and M.~Yamazaki,
\textit{``{Integrable Amplitude Deformations for N=4 Super Yang-Mills and ABJM
  Theory}''},
\textsf{\doiref{10.1103/PhysRevD.91.026004}{Phys.~Rev.~D91,~026004~(2015)}},
\texttt{\arxivref{1407.4449}{arxiv:1407.4449}}.

\bibitem{Frassek:2013xza}
R.~Frassek, N.~Kanning, Y.~Ko and M.~Staudacher,
\textit{``{Bethe Ansatz for Yangian Invariants: Towards Super Yang-Mills
  Scattering Amplitudes}''},
\textsf{\doiref{10.1016/j.nuclphysb.2014.03.015}{Nucl.~Phys.~B883,~373~(2014)}},
\texttt{\arxivref{1312.1693}{arxiv:1312.1693}}.

\bibitem{Kanning:2014maa}
N.~Kanning, T.~{\L}ukowski and M.~Staudacher,
\textit{``{A shortcut to general tree-level scattering amplitudes in
  $\mathcal{N} = 4$ SYM via integrability}''},
\textsf{\doiref{10.1002/prop.201400017}{Fortsch.Phys.~62,~556~(2014)}},
\texttt{\arxivref{1403.3382}{arxiv:1403.3382}}.

\end{thebibliography}
